\def\useplain{1}

\ifx \useplain\undefined
	\documentclass[preprint, nonblindrev]{informs3aa}
	\usepackage{informs_set}
	\input{informs_title}
\else
	\documentclass{article}
	\usepackage[margin=1.5in]{geometry}
	\usepackage{plain_macro}
	\title{Information and Memory in Dynamic Resource Allocation}

\author{
Kuang Xu\\
Graduate School of Business\\
Stanford University\\
   \texttt{kuangxu@stanford.edu} 
  \and
Yuan Zhong\\
Booth School of Business\\
University of Chicago\\
  \texttt{Yuan.Zhong@chicagobooth.edu} 
}

\date{}

\fi

\usepackage{kuang_macro}

\begin{document}

\ifx \useplain\undefined
	\input{informs_title}
\else
	\title{Information and Memory in Dynamic Resource Allocation}

\author{
Kuang Xu\\
Graduate School of Business\\
Stanford University\\
   \texttt{kuangxu@stanford.edu} 
  \and
Yuan Zhong\\
Booth School of Business\\
University of Chicago\\
  \texttt{Yuan.Zhong@chicagobooth.edu} 
}

\date{}

\fi

\ifx \useplain\undefined
\else
\maketitle
\fi

\def\red#1{{\color{red} #1}}

\def\abs_txt{
We propose a general framework, dubbed \emph{Stochastic Processing under Imperfect Information (SPII)}, to study the impact of information constraints and memories on dynamic resource allocation. The framework involves a Stochastic Processing Network (SPN) scheduling problem in which the {scheduler} may access the system state only through  a noisy {channel}, and resource allocation decisions must be carried out through the interaction between an encoding policy (who observes the state) and allocation policy (who chooses the allocation). Applications in the management of large-scale data centers and human-in-the-loop service systems are among our chief motivations.

We quantify the degree to which information constraints 
{reduce} the size of the \emph{capacity region} in general {SPNs}, and how such reduction depends on the amount of {memories} available to the encoding and allocation policies. Using a novel metric, {\emph{capacity factor}}, our main theorem  characterizes the reduction in capacity region (under ``optimal'' policies) for {all non-degenerate channels}, and across {almost all combinations of memory sizes}. Notably, the theorem demonstrates, in substantial generality, that $(1)$ the presence of a noisy channel {always reduces} capacity, $(2)$ more {memory} for the allocation policy {always improves} capacity,  and $(3)$ more {memory} for the encoding policy has little to no effect on capacity. Finally, all of our positive (achievability) results are {established through} constructive, implementable policies. 

Our proof program involves the development of a host of new techniques, largely from first principles, by combining ideas from information theory, learning and queueing theory.  As a sub-module of one of the policies proposed, we create a simple yet powerful generalization of the Max-Weight policy, in which individual Markov chains are selected dynamically, in a manner analogous to how schedules are used in a  conventional Max-Weight policy.\footnote{April 2019; revised August 2019. A preliminary version appears in the proceedings of {ACM SIGMETRICS} 2019 as an extended abstract. The Appendix can be found in the Supplemental Material. }

\emph{Keywords}: resource allocation, scheduling, max-weight algorithm, queueing, stochastic processing network, information theory, memory. 
}

\ifx \useplain\undefined
\ABSTRACT{\abs_txt}
\else
\begin{abstract}
\abs_txt
\end{abstract}
\fi


\ifx \useplain\undefined
\maketitle
\fi

\def\bB{\bm{B}}
\def\bQ{\bm{Q}}
\def\bD{\bm{D}}
\def\bE{\bm{E}}
\def\bA{\bm{A}}
\def\bM{\bm{M}}
\def\bW{\bm{W}}
\def\bR{\bm{R}}
\def\bS{\bm{S}}
\def\bZ{\bm{Z}}
\def\balpha{\bm{\alpha}}
\def\blambda{\bm{\lambda}}
\def\bmu{\bm{\mu}}
\def\bgamma{\bm{\gamma}}
\def\veps{\varepsilon}
\def\bd{\bm{d}}
\def\be{\bm{e}}
\def\bmm{\bm{m}}
\def\bq{\bm{q}}
\def\bw{\bm{w}}
\def\bx{\bm{x}}
\def\by{\bm{y}}
\def\bz{\bm{z}}
\def\conv{{\rm conv}}
\def\pconv{{\rm conv^-}}
\def\rel{{\rm rel}}
\def\bOne{\bm{1}}
\def\bZero{\bm{0}}
\def\cl{{\rm cl}}

\def\blam{\bm{\lambda}}
\def\bmu{\bm{\mu}}
\def\bD{\bm{D}}
\def\bLam{\bm{\Lambda}}

\def\kx#1{{\color{blue} #1}}

\def\rv#1{{\color{blue} #1}}

\def\red#1{{\color{red} #1}}

\section{Introduction}\label{sec:intro}

 In {many} modern large-scale resource allocation systems,  such as data centers, call centers and hospitals, getting reliable access to accurate system state information {usually requires} expensive investment in monitoring  or machine  learning infrastructures. Furthermore, such information can often be subject to noise, loss or misinterpretation.  It is therefore crucial to understand how \emph{imperfect} and {noisy} information {affects system performance}. 
Insights along this direction can provide crucial architectural guidelines on how to design efficient information-driven scheduling policies, and also assist {with} infrastructure planning by quantifying 
{the performance benefits from better information access, 
and hence the tradeoffs between such benefits and the investment costs.}
 
 As a first step towards this direction, we propose in this paper a new, general framework {for quantifying} the performance impact of  information constraints on an underlying dynamic resource allocation problem. Specifically, we will focus on Stochastic Processing Networks (SPN) \citep{harrison2002stochastic, Harrison2000spn, Harrison2003correction, DaiLin2005}
{-- a widely used paradigm for modeling resource allocation problems in diverse sectors, 
including information technology \citep{DaiPrabhakar2000, McKeown1996, RobertsMassoulie2000, tassiulas1992stability}, 
manufacturing \citep{Wafer}, call centers \citep{harrison2005method, Gans2003, mandelbaum2004scheduling}, as well as other service industries --} 
and address how information, or the lack of, alters the capacity region of an SPN. 
 
\begin{figure}
\centering
\includegraphics[scale=.55]{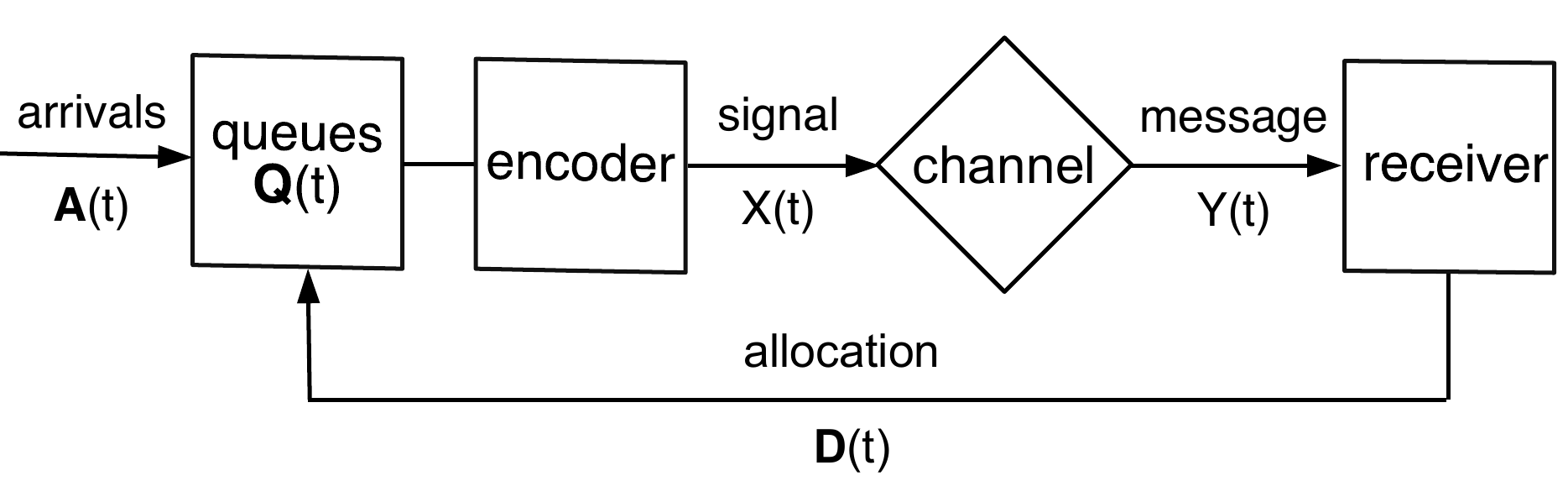}
\caption{\normalsize System diagram. The encoder is able to fully observe the state of the queues, $Q(t)$. The encoder and receiver are equipped with memories of size $k$ and $v$ bits, respectively. } 
\label{fig:sysDiag}
\end{figure}

We begin with an overview of the model; the formal description will be presented in Section \ref{sec:Model}.  The framework, dubbed the {\bf Stochastic Processing under Imperfect Information} (SPII) model, is illustrated in Figure \ref{fig:sysDiag}, and consists of three main elements: an underlying dynamic resource allocation problem, a model of imperfect information, and memories.

\noindent {\bf I.~Stochastic Processing Network Scheduling}. The underlying dynamic resource allocation problem is that of scheduling in a discrete-time SPN, sometimes also known as a \emph{switched network} \citep{stolyar2004maxweight, shah2012switched}, where a finite set of processing resources is employed to serve incoming tasks of $N$ different \emph{types}. In each time {slot} $t$, new tasks arrive to the system in a stochastic manner, where unprocessed tasks of type $i$ are buffered in a queue $i$, and the queue lengths are denoted by $\bQ(t) = (Q_1(t), \ldots, Q_N(t))$.  The number of arrivals of type $i$ jobs {at time $t$}, $A_i(t)$, has an expected value $\E[A_i(t)]=\lambda_i$, and we  refer to the vector $\blam = (\lambda_1, \ldots, \lambda_N)$ as the \emph{arrival rate vector}.  The \emph{scheduler} is to select from a finite \emph{schedule set}, $\Pi$, an \emph{allocation vector}, $\bD(t) = (D_1(t), \ldots, D_N(t)) \in \Pi$, where $D_i(t)$ corresponds to the number of tasks in queue $i$ that can be processed during the present time {slot}. The queue lengths evolve according to the following dynamics\footnote{{For a vector $\bx = (x_1, \ldots, x_N)$, 
we use the notation $(\bx)^+$ to mean the vector $(\max\{x_1, 0\}, \ldots, \max\{x_N, 0\})$.}}: 
\begin{equation}
\bQ(t) = \left(\bQ(t-1)-\bD(t)\right)^+ + \bA(t), \quad t\in \N.
\end{equation}

\begin{figure}
\centering
\includegraphics[scale=.55]{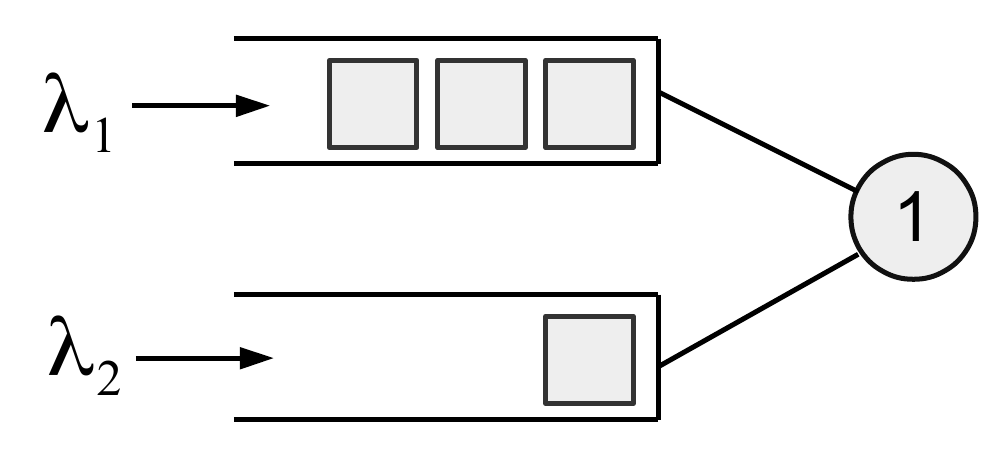}
\caption{An example of a Stochastic Processing Network with one server and two parallel queues. Two types of tasks arrive to the systems  at rates $\lambda_1$ and $\lambda_2$ respectively, and wait in two  separate queues. The schedule set consist of two possible allocation vectors, $(1,0)$ and $(0,1)$, representing the server's choice of processing one job from queue 1 or queue 2, respectively. }
\label{fig:twoqueues}
\end{figure}

As an example, the scheduling problem involving one server and two parallel queues falls under this framework, as illustrated in Figure \ref{fig:twoqueues}, where the schedule set contains two elements, $\Pi = \{ (1,0),(0,1) \}$, corresponding to the server processing a job from the first and second queue, respectively. 

\noindent {\bf II.~Scheduling with Imperfect Information}. 
{Here, we discuss} the information that is available to the decision maker in our model. 
{As we will explain, the decision maker consists of two separate, though coordinating sub-entities, 
the \emph{encoder} and the \emph{receiver}, respectively; 
the former can observe the queue lengths, while the latter chooses the allocations. 
This is very different from prior literature on SPNs, where the decision maker is synonymous with the scheduler. 
In contrast, under our framework, the receiver is the scheduler; the term ``receiver'' is used to emphasize the fact that 
allocation decisions are made based on messages received through an information channel.} 

{We now proceed to describe the information structure in our model in detail.}
First, similar to much of the prior literature, we assume that the arrival rate vector $\blam$ is not known to the decision maker; 
otherwise, she can simply choose a stationary, randomized allocation vector that dominates $\blam$ to stabilize the system. Second, as a major departure from conventional SPNs, the {scheduler} in our model {does not} have access to the full queue-length state information when making {allocation} decisions. In contrast, she obtains information concerning the queues only through a {\bf (noisy) channel}. A channel consists of a pair of finite input and output alphabets, $\calX$ and $\calY$, and a family of probability distributions $\{\pb^C_x\}_{x\in \calX}$ over $\calY$. When an input \emph{signal}, $x \in \calX$, is sent through the channel, it results in a random output \emph{message}, $Y \in \calY$, drawn from the probability distribution $\pb^C_x(Y = \cdot)$. For instance, one simple channel is that of an $\veps$-noisy binary symmetric channel with $\calX=\calY=\{0, 1\}$, where the input signal is correctly received with probability $1-\veps$, and is perturbed to the opposite symbol with probability $\veps$, i.e., $\pb^C_{a}(Y=a) = 1-\veps$ and $\pb^C_a(Y=b) = \veps$ for $a,b \in \{0,1\}$, $a\neq b$. By using different alphabets and distributions $\pb^C$, the channel is able to capture a variety of partial {and/or lossy} information models, such as controlling a data center over band-limited communication constraints, where only noisy {and/or} compressed signals of the full system {state} can be obtained. In this paper, we will impose little 
restriction on the form of the channel, in order to allow for a maximum degree of generality and modeling flexibility. 

In the presence of a channel, the allocation decisions in our model are carried out by a pair of {\bf encoding} and {\bf allocation policies}, situated on opposite ends of the channel (Figure \ref{fig:sysDiag}). An \emph{encoder} is co-located with the queues and has complete knowledge of the queue lengths at all times. In time {slot} $t$, the encoder employs an \emph{encoding policy}, $\phi$, to send an \emph{input symbol} $X(t)\in \calX$ through the channel. The resulting random \emph{output message}, $Y(t)$, arrives at the \emph{receiver}\footnote{{As explained earlier, the receiver is the same as the scheduler in our model. For the rest of the paper, we will use the term ``receiver'' for consistency of terminology.}}, who then employs an \emph{allocation policy}, $\psi$, to choose the allocation vector $\bD(t)$. Note that the receiver does not observe the queue lengths, and in most settings {considered in this paper}, the information contained in $Y(t)$ is severely limited; for instance, the output alphabet $\calY$ can be substantially smaller than the number of possible allocation vectors in $\Pi$. 

\noindent {\bf III.~Memory.} The last crucial element of the system is memory. In a classical SPN, {external memory is} rarely needed by a scheduling policy, because the queue lengths are fully observable and already contain all the relevant information for policies such as Max-Weight \citep[e.g.,][]{tassiulas1992stability, DaiLin2005, stolyar2004maxweight, shah2012switched} to make scheduling decisions. However, when the system state information becomes constrained and imperfect, memory {plays} a crucial role in determining the system performance 
{(for details, see Theorem \ref{thm:main} and the subsequent remark)}. For instance, with memory, the allocation policy could aggregate past messages across multiple {time slots} 
to better estimate the system state and inform its scheduling decisions, and similarly, the encoding policy could benefit by remembering past transmissions in order to better tailor future input signals.

For this reason, we will allow for the possibility that the encoder and receiver have at their disposal a finite \emph{memory} of {$k$ and $v$ bits}, respectively, in which  information can be recorded and subsequently retrieved in the next time {slot}. The encoding policy may generate the input signal $X(t)$ based on the queue lengths as well as the state of the encoder memory, and similarly, the allocation policy may choose the allocation vector $\bD(t)$ based on the received output message $Y(t)$ along with the state of the receiver memory. If $k$ or $v$ is equal to 0, we say that the corresponding encoding or allocation policy is \emph{memoryless}. 

Finally, we say that the system has {\bf memory-feedback} if the state of the receiver memory is accessible by the encoding policy. The existence of  memory-feedback  would correspond to applications in which the \emph{backward} communication from the receiver to the encoder is {lossless and} relatively cheap. Arguably, this is not a very restrictive assumption within the SPII framework,  considering that the receiver must, in any event, communicate the allocation decision, $\bD(t)$, back to the queues, and hence sending along the state of the receiver memory in the mean time should not incur too much additional overhead. A number of main results in the present paper will rely on this assumption, and we expect to relax it in future work. 

\begin{figure}[h]
\centering
\includegraphics[scale=.5]{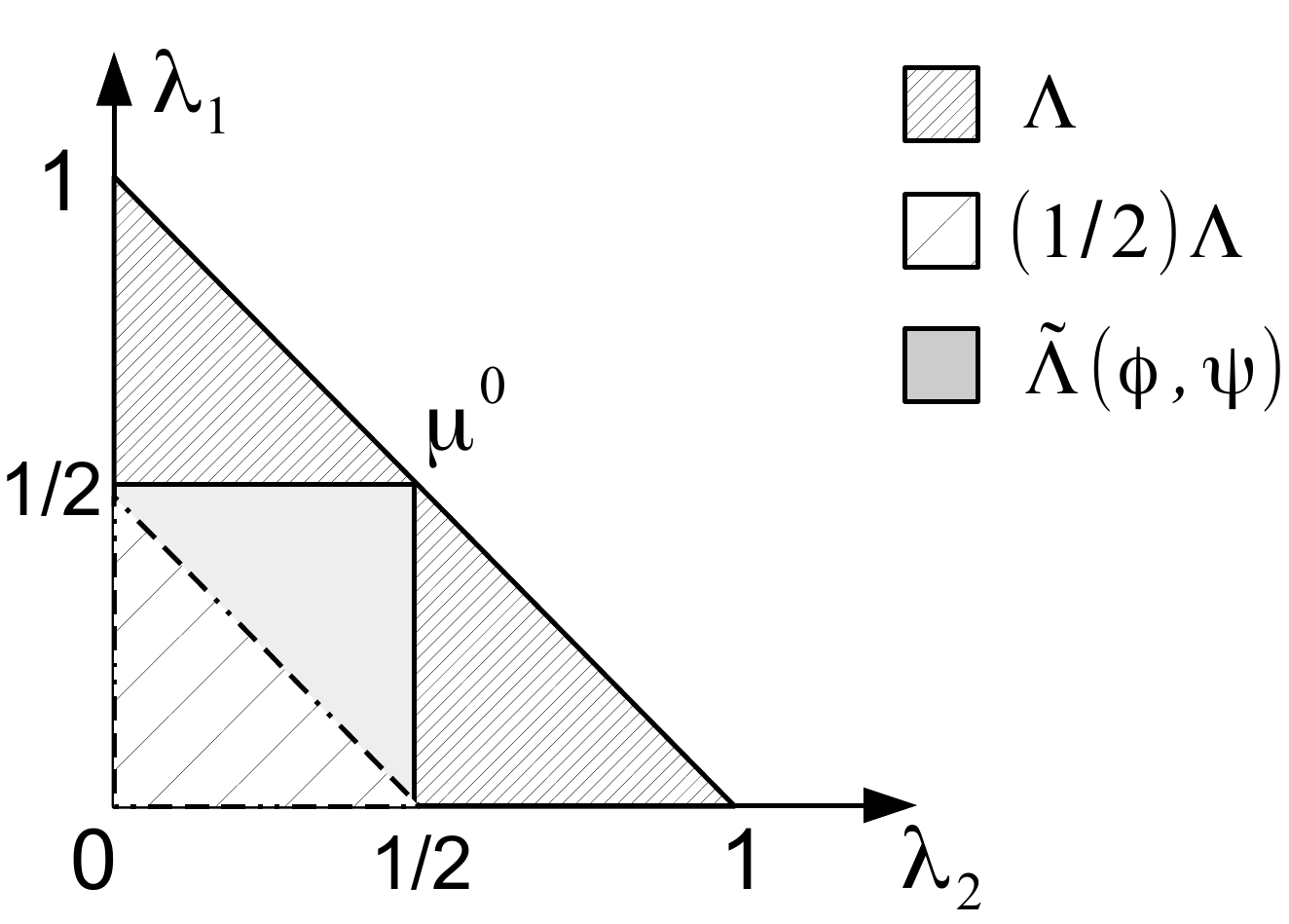}
\caption{Illustration of the reduction of capacity region due to lossy information, for the two-queue-one-server SPN in Figure \ref{fig:twoqueues}. The triangle with vertices $(0,0)$, $(0,1)$ and $(1,0)$ represents the maximum capacity region, $\bLam$, when the queue lengths are fully observable. The rectangle dominated by $\bmu^0$ represents the capacity region $\tilde{\bLam}(\phi, \psi)$ under an uninformative channel and {a memoryless} allocation policy that induces an average service rate of $\bmu^0$. The triangle with vertices $(0,0)$, $(0, 1/2)$ and $(1/2,0)$ represent the scaled (``shrunk'') version of the maximal capacity region that is contained in $\tilde{\bLam}(\phi, \psi)$.}
\label{fig:capLoss1}
\end{figure}

\subsection{Performance Metric: Capacity Factor} 

The main quantity of interest is the \emph{capacity region} of an SPII: the set of arrival rate vectors $\blam$ under which the queue lengths, $\bQ(\cdot)$, remain stable\footnote{{Roughly speaking, the process $\bQ(\cdot)$ is stable if queue lengths do not grow to infinity as time goes. The precise notion of stability that we consider in this paper will be defined in Section \ref{ssec:cap_fac}.}}, for a given pair of encoding and allocation policies. As we shall see shortly, harsh information constraints tend to {reduce} the capacity region. 
{Therefore, at a high level, the main question that we aim to address can be stated as follows:}

\emph{How does the {information constraint}, in the form of a noisy channel, reduce the capacity region of an SPII architecture, and how does the degree of reduction  vary with respect to the sizes of encoder and receiver {memories}? 
}

In order to {measure} the {magnitude} of the reduction in capacity region, we introduce a key (scalar) metric, dubbed the {\bf capacity factor},  which captures the {fraction} of capacity region lost due to imperfect information. 
{Informally, the notion of capacity factor can be described as follows (the formal definition will be given in Section \ref{sec:Model}).}
Fix a channel $\calC = (\calX, \calY, \{\pb^C_x\})$. {For a pair of encoding-allocation policies $(\phi, \psi)$}, denote by $\tilde{\bLam}(\phi, \psi)$ the set of all arrival rate vectors $\blam$ that the system is able to stabilize under the pair $(\phi, \psi)$ and channel $\calC$. Denote by $\bLam$ the \emph{maximum capacity region} of the SPII, defined by the set of arrival rate vectors that are dominated by the convex hull of the schedule set, $\Pi$. In particular, $\bLam$ corresponds to the maximal capacity region of the system under \emph{full information} and a maximally stable scheduling policy, such as Max-Weight. 
\begin{definition}[Capacity Factor (Informal)] Fix $k,v \in \zp$. The {$(k, v)$-\emph{capacity factor}} of the channel $\calC$, denoted by $\rho^*_{k,v}(\calC)$, is defined by the {supremum} of $\rho \in (0,1)$, such that 
\begin{equation}
\rho \bLam \subset \tilde{\bLam}(\phi, \psi), 
\end{equation}
{for {some} pair of encoding and allocation policies $(\phi, \psi)$, 
whose memory sizes are $k$ and $v$ bits, respectively}.
\end{definition}

{A major motivation for introducing the notion of capacity factor is for analytical tractability. 
In general, for large-scale systems and when the encoder and/or the receiver has a finite-sized memory, 
the set $\tilde{\bLam}(\phi, \psi)$ can have very complicated geometry (see Example 1 below for an illustration), 
making it hard to use for performance characterization directly. Instead, the capacity factor is a simple scalar performance metric that is much more tractable, even though it does not capture the entire geometry of the set $\tilde{\bLam}(\phi, \psi)$.}
{Moreover, the capacity factor admits an intuitive interpretation as the largest fraction of the full-information maximum capacity region that can be ``preserved'' under information constraints (using an ``optimal'' encoding-allocation policy pair).} A capacity factor of $1$ means that the information constraints result in {no loss} of the maximum capacity region, and on the other extreme, a capacity factor of $0$ indicates that {most} of the maximum capacity region has been lost due to {lossy} information. 

\begin{example}[Scheduling with Parallel Queues]
\label{ex:paraQ}
We now present a simple example to illustrate 
{the drastic impact that lossy information can have on the capacity region}, 
consequently resulting in a small capacity factor. Consider the SPN in Figure \ref{fig:twoqueues} with one server and two parallel queues. It is not difficult to see that the maximum capacity region under full information, $\bLam$, consists of all arrival rates where $\lambda_1+\lambda_2 <1$,  achieved by always serving a queue that is non-empty.

Now suppose  the channel of the SPII is completely uninformative, so that the output message $Y(t)$ is independent of the input signal $X(t)$. 
{Furthermore, for simplicity, let us restrict our attention to memoryless allocation policies.}
Then, {since the allocation decisions can only depend on the current, uninformative, output message $Y(t)$, 
the allocation policy will choose to serve an empty queue with positive probability. Indeed,} 
it can be shown that any allocation policy will induce a {fixed} average service rate vector, $\bmu^0$, independent of the actual arrival rates. 
The resulting capacity region is represented by the rectangle in Figure \ref{fig:capLoss1} (dominated by $\bmu^0$), which is  {always} substantially smaller than the maximum capacity region, regardless of the choice of $\bmu^0$.  It is not difficult to show that the {capacity factor} under such an uninformative channel is equal to $1/2$, which corresponds to the case where the allocation policy induces a service rate vector of $\bmu^0=(1/2,1/2)$ by choosing the two schedules with equal frequency. 
Therefore, depending on the quality of the channel and the memories available, the fractional loss of capacity region in SPII can range from $1/2$ (uninformative) to $1$ (full-information). 
\end{example}

\begin{table}[h]
\renewcommand{\arraystretch}{1.5}
\begin{center}
\begin{tabular}{ |c | c | c| c| }
\hline 
Values of  $\rho^*_{k,v}(\calC)$& $k=0 $ & $k \in \N$  & $k= \infty$ \\ [0.5ex] 
\hline
 $v=0$ & $\rho^*_{0,0}=\rho^*_{\infty,0} $ & $\hspace{-.5cm}{^{\Rightarrow}}~\rho^*_{k,0}=\rho^*_{\infty,0}$ & $\rho^*_{\infty,0}<1 {\,}^{\dagger}$  \\  
\hline
  $v \in \N$ & $-$ &$\rho^*_{K(\Pi,\calX), v}=\rho^*_{\infty, v} $ & $ \rho^*_{\infty, v}<1 {\,}^{\dagger}$   \\
 \hline
  $v = \infty$ & $1$ & $\hspace{-.5cm}{^{\Rightarrow}}~1$  & $\hspace{-.5cm}{{\,}^{\Rightarrow}}~1$   \\
 \hline
\end{tabular}
\end{center}
\vspace{20pt}
\caption{Summary of the results in Theorem \ref{thm:main}, where we omit the dependence on $\calC$ for simplicity of notation. The system is assumed to have memory-feedback. The parameters $k$ and $v$ denote the size of the memory for the encoder and receiver, respectively.
{The symbol ``${\,}^{\Rightarrow}$''} indicates that the result of the cell is readily implied by that of the cell to the left. The symbol ``${\,}^\dagger$'' means that the result holds if in addition to being informative, the channel is also ``noisy'' (more precisely, $\veps$-majorizing for some $\veps>0$; see Definition \ref{def:eps-Major}). Finally, the symbol ``$-$'' means that we do not yet know the exact value of $\rho_{k,v}^*(\calC)$ when $k<K(\Pi, \calX)$, beyond the fact that it's not greater than $\rho^*_{K(\Pi,\calX), v}(\calC)$. For instance, we do not know whether $\rho_{k,v}^*(\calC)$ is in fact equal to $\rho^*_{K(\Pi,\calX), v}(\calC)$ for all $k<\rho_{k,v}^*(\calC)$.}
\label{tab:main}
\end{table}

\subsection{Preview of Main Results: Characterization of Capacity Factor} 
We now give an informal preview of our main results (Theorem \ref{thm:main}), which are also summarized in Table \ref{tab:main}. Fix any channel that is \emph{informative}, such that the input signal $X(t)$ and output message $Y(t)$ are {not} independent. Suppose that the system {has memory-feedback}. We show the following: 
\begin{enumerate}
\item When the receiver is memoryless ($v =0$), the capacity factor $\rho^*_{k,0}(\calC)$ is
independent of $k$, {the size of the encoder memory}: 
\begin{equation}
\rho^*_{k,0}(\calC) = \rho^*_{0,0}(\calC), \quad \forall k \geq 0. 
\end{equation}
\item When the receiver has a finite memory ($0<v<\infty$), the capacity factor $\rho^*_{k,v}(\calC)$ may depend on $k$, but only up to a constant, $K(\Pi, \calX)$, whose value does not depend on $v$ or the channel $\calC$:
\begin{equation}
\rho^*_{k,v}(\calC) =\rho^*_{K(\Pi, \calX),v}(\calC), \quad \forall  k  \geq K(\Pi, \calX). 
\end{equation}
\item In the limit where {the size of the receiver memory} tends to infinity ($v=\infty$), the (limiting) capacity factor $\rho^*_{k,\infty}(\calC)$ is always $1$, regardless of {the magnitude of $k$}:  
\begin{equation}
\rho^*_{k,\infty}(\calC) =1, \quad \forall  k  \geq 0. 
\end{equation}
In particular, there will be  {no loss} of capacity region in this limiting regime. 
\item Finally, if the channel is ``{noisy}'' (to be precisely defined), and the receiver does not have infinite memory ($v <\infty$), then the capacity factor $\rho^*_{k,v}(\calC)$ is {always} strictly less than 1:
\begin{equation}
\rho^*_{k,v}(\calC) <1, \quad \forall v <\infty.
\end{equation}
\end{enumerate}

Moreover, as a by-product of the proof,  all of our achievability results (Items 1 through 3) are {established through} {constructive} and implementable encoding-allocation policy pairs. The capacity factor can also be {explicitly computed} in closed-form for restricted families of channels and schedule sets, which we will demonstrate in Section \ref{sec:epsMaj}. {Appendix \ref{app:calcRho} discusses how to compute the capacity factor and identify optimal policies that achieve the capacity factor for the general case.}

{\bf Key implications of the main results}: Our results have a number of architectural implications; we highlight some of them below.
\begin{enumerate}
\item \emph{Impact of imperfect information is non-trivial}. The result in Item 4 shows that the presence of a noisy channel {always} reduces capacity (i.e., $\rho^*_{k,v}(\calC) <1$), regardless of the amount of  memories available to the encoding or allocation policies, so long as the allocation policy does not have infinite memory. 
\item  \emph{Memories play an intricate, asymmetric role}.  The results in Items 1 through 3 demonstrate that {memories} are crucial in mitigating the reduction in capacity region due to imperfect information.   However, memories are {substantially more important} for the allocation policy than they are for the encoding policy:  increase in memory for the allocation policy {always} improves capacity, while the benefit of having additional memory for the encoding policy becomes {zero} beyond some finite threshold. This asymmetry suggests design principles that favor simple, low-memory communication modules for the encoding policy, while the allocation policy must aggregate sufficient amount of past observations in order to achieve a maximum capacity region.
\end{enumerate}

\subsubsection{Motivating Examples of SPII}
 While the SPII model is highly stylized, it is motivated by a range of dynamic resource allocation systems with information constraints. We discuss in this sub-section {two} motivating examples: 
 
 \begin{enumerate}
 \item \emph{Example 1 - Scheduling in Large-Scale Data Centers with Limited Communication.} {It has long been recognized that obtaining reliable system-wide state information 
 is challenging in modern-day data centers, where many tasks are processed in a massively parallel manner, 
 servers may fail, and bandwidths may be limited  \cite[e.g.,][]{Armbrust2010, Mosharaf2015, Fastlane2015, Grass2014}.
 {For a concrete illustration,} 
 consider a manager {(the receiver that decides on the allocation policy)} who is operating a large-scale data center via a limited communication channel, 
{through which the data-center servers (the encoder) transmit information about their states.}
Here, the channel is broadly construed as incorporating all aspects of information obfuscations, such as those due to server failures, and limiting bandwidths on the network and server I/O. {For instance, because of communication constraints on the network and server I/O, the overall state of the servers may need to be condensed and sent to the manager along a rate-limited communication link. The channel model in the SPII framework can capture this scenario: suppose that the communication link is capable of noiselessly transmitting $m$ bits of information per time slot. Then, the channel would be one in which the input and output alphabets are the set of all $m$-bit strings, $\{0,1\}^m$,  and the channel matrix a $2^m \times 2^m$ identity matrix. In other examples, the channel can also be used to model dropped packets or noisy measurements of the overall system state. }  Overall, our model will speak to the design of the communication protocol from the data center to the manager in these scenarios, as well as how  she should translate the received messages into resource allocation decisions.}
 \item \emph{Example 2 - Human-in-the-Loop Resource Allocation.}  Our model can also be applied to resource allocation problems where communications amongst human operators may be subject to errors or misinterpretations \citep[e.g.,][]{manojlovich2015effect, nursehandover, traumahandover}. For instance, one may consider a hospital setting where  {nurses} (``encoder''), who observe the state of the ward, must communicate the relevant information and instructions to {the physician} (``receiver''), who will carry out the actual interventions. {Misunderstanding or loss could occur when transmitting information from the nurses to the physicians leading to misplaced decisions. Because both the encoder and receiver in this system can be human agents, memories can be a highly constraint resource and it is thus important to understand how to design simple and low-memory policies that may be just as effective at delivering optimal system capacity. }

 \end{enumerate}

\subsection{Organization}  
The remainder of the paper is organized as follows. We formally describe our model in Section \ref{sec:Model}, and present our main results in Section \ref{sec:main_res}. Section \ref{sec:lit_review} discusses the related literature. The remainder of the paper is devoted to the proof of our main results, with a proof overview in Section \ref{sec:poverview} that summarizes the key techniques. {We conclude the paper with some discussion in Section \ref{sec:conclude}}.

\subsection{Notation}\label{ssec:notation}
We reserve boldface letters for vectors and plain letters for scalars. 
For any scalar $x$, $\lceil x \rceil$ denotes the smallest integer greater than or equal to $x$, 
and $(x)^+ \triangleq \max\{x, 0\}$ denotes the non-negative part of $x$. 
For any $d\in \N$, and vector $\bx \in \R^d$, we use $x_i$ to denote the $i$th coordinate of $\bx$. 
We use $(\bx)^+$ to denote the vector $\lt((x_1)^+, \ldots, (x_d)^+\rt)$.
For any $\bx,\by\in \rp^d$, 
we write $\bx\leq \by$ if $x_i \leq y_i$ for all $i=1, 2, \ldots, d$, 
and we write $\bx < \by$ if $x_i < y_i$ for all $i=1, 2, \ldots, d$. For a set $\calX \subset \rp^d$, we write $\bx\leq \calX$ ($\bx < \calX$, respectively) if there exists $\by \in \calX$ such that $\bx\leq \by$ ($\bx < \by$, respectively), and say that the vector $\bx$ is {dominated} ({strictly dominated}, respectively)  by the set $\calX$. 
{We call a vector $\bz$ a {\em maximal element} of the set $\calX \subset \rp^d$ if (a) $\bz \in \calX$; 
and (b) for any $\by \in \calX$, $\bz \leq \by$ implies $\bz = \by$.}
For $N\in \N$, we will use the short-hand $[N]$ to denote the set $\{1, 2, \ldots, N\}$ of consecutive integers. 

For a set $\calS$ and function $f$, we will use $f(\calS)$ to denote the set $\{f(s): s\in \calS\}$. 
For any set $\calS \in \rp^d$, we use $\conv(\calS)$ to denote the convex hull of $\calS$, 
$\pconv(\calS)\bydef \{\bx \in \rp^d: \bx < \conv(\calS)\}$ to denote 
the set of vectors $\bx \in \rp^d$ {strictly} dominated by $\conv(\calS)$, 
and $\cl(\calS)$ to denote the closure of the set $\calS$.

For vectors $\bx, \by \in \R^d$, the inner product of $\bx$ and $\by$ is denoted by 
$\langle \bx, \by\rangle \triangleq \sum_{i=1}^d x_i y_i$. The $v_2$-norm 
of vector $\bx$ is denoted by $\|\bx\| \triangleq \sqrt{\sum_{i=1}^d x_i^2}$, 
and the $v_{\infty}$-norm denoted by $\|\bx\|_{\infty} \triangleq \max\{|x_1|, \ldots, |x_d|\}$. 
The $i$th standard unit vector in $\R^d$ is denoted by $\be^{(i)}$, 
whose $i$th coordinate equals $1$ and all other coordinates equal zero. 
The vector with all components being $1$ is denoted by $\bOne$.

We use the shorthand ``w.p.1'' to mean ``with probability 1,'' 
``i.i.d'' to mean ``identically and independently distributed,'' and 
``WLOG'' to mean ``without loss of generality.''  We use $\text{Uniform}(0, 1)$ to describe random variables that are uniformly distributed over 
the interval $(0, 1)$. The indictor function of an event $A$ is denoted by $\mathbb{I}_A$.

\section{The Model}
\label{sec:Model}

We formally present the model, Stochastic Processing with Imperfect Information (SPII), in this section. 

{\bf Stochastic Processing Network model.} We consider a dynamic Stochastic Processing Network (SPN) evolving in discrete time $t\in \zp$ (Figure \ref{fig:sysDiag}). The system consists of $N$ queues, whose lengths at the end of the $t^{\text{th}}$ time slot are represented by $\bQ(t) \in \zp^{N}$. The evolution of the queues is captured by the following equation:
\begin{equation}\label{eq:dynamics}
\bQ(t) = \left(\bQ(t-1)-\bD(t)\right)^+ + \bA(t), \quad t\in \N,
\end{equation}
where $\bA(t)$ and $\bD(t)$ are the arrival and allocation vectors during slot $t$, respectively. We use $\Pi \subset \zp^{N}$ to denote the finite set of all allowable schedules. 
{For each $i \in [N]$, the arrivals over time} are i.i.d.~Bernoulli random variables 
{that are independent from everything else}, where 
\begin{equation}
\E(A_i(1)) = \lambda_i, \quad i =1, 2, \ldots, N. 
\end{equation}
We will refer to $\blambda = (\lambda_1, \lambda_2, \ldots, \lambda_N)$ as the \emph{arrival rate vector}.

WLOG we suppose that the schedule set 
$\Pi$ satisfies the following three assumptions, {throughout the paper}. 
\begin{asmp}\label{asmp:monotone}
$\Pi$ is {\em monotone}: if $\bd \in \Pi$, 
then for any $\bd' \in \zp^N$ with $\bd' \leq \bd$, 
$\bd' \in \Pi$ as well.
\end{asmp}
\begin{asmp}\label{asmp:non-degeneracy}
{There exists some $c>0$ such that for each $i \in [N]$, $c\be^{(i)} \in \Pi$.}
\end{asmp}
\begin{asmp}\label{asmp:non-degeneracy2}
The schedule set $\Pi$ has at least two distinct maximal elements. 
More specifically, there exist $\bd^{(1)}, \bd^{(2)} \in \Pi$ 
with $\bd^{(1)} \neq \bd^{(2)}$, and for $j = 1, 2$, 
$\bd^{(j)} \leq \bd$ with $\bd \in \Pi$ implies that $\bd = \bd^{(j)}$.
\end{asmp}
Assumption \ref{asmp:monotone} allows for more flexible allocation decisions without impacting system performance. 
Assumption \ref{asmp:non-degeneracy} guarantees that each queue can receive a positive service rate, 
and Assumption \ref{asmp:non-degeneracy2} rules out the possibility of 
$\Pi$ having only one maximal element, in which case the single maximal element would dominate all other schedules 
{(see Lemma \ref{lem:max_extreme1} of Appendix \ref{app:thm:kleps<1} for details)}, 
and the trivial decision of always choosing that maximal element would be optimal 
for a wide range of performance objectives, such as maximizing throughput or minimizing queue lengths.


{\bf Signals, channels, and messages.}  A \emph{channel} is a {triplet} $\calC = (\calX, \calY, C)$, where $\calX$ and $\calY$ are finite sets representing the input and output alphabets, with cardinalities $\szX$ and $\szY$, respectively, and $C$ is an $\szX$-by-$\szY$ row stochastic matrix, which we will refer to as the \emph{channel matrix}. Since $C$ is row stochastic, each row corresponds to a probability distribution over $\calY$, the set of output alphabets, and we denote the probability distribution corresponding to the $x$th row as $\pb^C_x$. When an input \emph{signal}, $X \in \calX$ is sent through the channel, it leads to a (possibly random) output \emph{message}, $Y \in \calY$, drawn from the probability distribution $\pb^C_x(Y = \cdot)$.
Thus, the matrix $C$ captures the stochastic distortion introduced by the channel, where the entry  $C_{x,y}$ represents the probability that the output message of the channel is $y$ when the input signal is $x$: 
\begin{equation}
C_{x,y} = \pb^C_x(Y = y), \quad x \in \calX, y\in \calY.
\end{equation}
We assume the channel is memoryless, so that each output message only depends on the input signal of the present time slot, and is independent from the system's past history. 
We also assume that the channel is stationary, so that for any $x\in \calX$ and $y \in \calY$, 
the probability that the output is $y$ when the input is $x$ does not depend on time.

{\bf Encoder and encoding policies.} During each time slot, an \emph{encoder} situated at the queues sends a {signal}, $X(t) \in \calX$, over the channel. The encoder is equipped with a finite-sized lossless \emph{memory} represented by a $k$-bit binary sequence, whose value at time $t$ is denoted by $M_e(t)$. The {signal} $X(t)$ can depend on the most recent state of the queues $\bQ(t-1)$, {the most recent arrivals $\bA(t-1)$}, 
the content of the {memory} $M_e(t-1)$, and possibly some idiosyncratic randomness. 
Formally, let $\phi_e$ be a deterministic \emph{encoding policy}. Then,\footnote{
In the sequel, we will see that our achievability results for the case of finite receiver memory 
(Items 1 and 2 of Theorem \ref{thm:main}; also Sections \ref{sec:memoryless} and \ref{sec:finRecMem}) 
are established using encoding policies that do not depend on $\bA(t-1)$, the most recent arrivals. 
However, to prove the result on infinite receiver memory (Item 3 of Theorem \ref{thm:main}; also Appendix \ref{app:sec:infRecMem}), the encoding policy that we constructed makes crucial use of $\bA(t-1)$, 
which is therefore included in Eq.~\eqref{eq:encoding}.}

\begin{equation}\label{eq:X_update}
X(t) = \phi_e(\bQ(t-1), \bA(t-1), M_e(t-1), U_e(t-1)), \quad t\in \N, 
\end{equation}
where $\{U_e(t)\}_{t\in \zp}$ is an string of i.i.d.~$\text{Uniform}(0, 1)$ random variables, which are all independent from the rest of the system. In each time slot $t$, the content of the memory, $M_e(t)$, is also updated based on $\bQ(t-1)$, {$\bA(t-1)$}, $M_e(t-1)$, and $U_e(t-1)$, and we can formally write
\begin{equation}\label{eq:Me_update}
M_e(t) = \phi_m(\bQ(t-1), \bA(t-1), M_e(t-1), U_e(t-1)), \quad t\in \N,
\end{equation}
for some deterministic function $\phi_m$. If we write $\phi = (\phi_e, \phi_m)$, 
then 
\begin{equation}\label{eq:encoding}
(X(t), M_e(t)) = \phi(\bQ(t-1), \bA(t-1), M_e(t-1), U_e(t-1)), \quad t\in \N. 
\end{equation}
With a slight abuse of notation, we also call $\phi$ the {\em encoding policy}. 
We will denote by $\Phi_k$ the set of all encoding policies with $k$ bits of memory. 
We use $\calM_e(k)$ to denote the set of possible values for the encoder memory when $k$ bits are allowed. 
When the context is clear, we often suppress the dependence on $k$ and simply write $\calM_e$.

Roughly speaking, the size of the memory, $k$, serves as a measure of ``complexity'' of an encoding function. A special case is when $k=0$, where the encoder is equipped with no memory and the signal depends only on the current state of the queues. We will refer to a policy $\phi\in \Phi_0$ as a \emph{memoryless encoding policy}. 

{\bf Receiver and allocation policies.} The signal $X(t)$ passes through the channel and results in a {message}, $Y(t) \in \calY$ at the \emph{receiver}. The responsibility of the receiver is to choose, at each time slot, the allocation vector $\bD(t)$. However, the receiver is not able to observe the state of the queues directly, so the allocation decisions can only rely on information provided by the encoder through the channel.  Similar to the encoder, the receiver is equipped with a memory of $v$ bits, whose state in slot $t$ is denoted by $M_r(t)$. Let $\psi_a$ be a deterministic \emph{allocation policy}, such that
\begin{equation}\label{eq:D_update}
\bD(t) = \psi_a(Y(t), M_r(t-1), U_r(t-1)),  \quad t\in \N, 
\end{equation}
where $\{U_r(t)\}_{t\in \zp}$ is an string of i.i.d.~$\text{Uniform}(0, 1)$ random variables, which are all independent from the rest of the system. In each time slot $t$, the content of the receiver memory, $M_r(t)$, is also updated. 
However, different from the allocation decisions, $M_r(t)$ is updated with a {time lag}, 
and it depends on $M_r(t-1)$, $U_r(t-1)$ and $Y(t-1)$, the message from an earlier time slot, instead of $Y(t)$, the most recent message.\footnote{{The one-step lag can be removed without substantially changing the results; it serves the purpose of simplifying the notation and proof.}} Formally, 
\begin{equation}\label{eq:Mr_update}
M_r(t) = \psi_m(Y(t-1), M_r(t-1), U_r(t-1)), \quad t\in \N,
\end{equation}
for some deterministic function $\psi_m$. Similar to the encoder side, 
we also call $\psi \bydef (\psi_a, \psi_m)$ the {\em allocation policy}. 

We will denote by $\Psi_{v}$ the set of all allocation policies with $v$ bits of memory. Analogous to the encoding policies, an allocation policy with no memory generates the allocation decision using only the current message, $Y(t)$. We will refer to a policy $\psi\in \Psi_0$ as a \emph{memoryless allocation policy}. The set of possible values for the receiver memory is denoted by $\calM_r(v)$ when $v$ bits are allowed. Similar to $\calM_e$, 
when the context is clear, we suppress dependence of $\calM_r(v)$ on $v$ and simply write $\calM_r$.

In this paper, we are primarily interested in the dynamics of the tuple 
\begin{equation}\label{eq:W}
\bW(t) \bydef \left(\bQ(t), \bA(t), M_e(t), X(t), M_r(t)\right), \quad t \in \zp.
\end{equation}
It is not difficult to verify that under any well-defined encoding-allocation policy pair $(\phi, \psi)$, 
$\bW(\cdot)$ is a countable-state Markov chain.
Therefore, for any time $t$, 
we call $\bW(t)$ \emph{the system state at time $t$}, and from now on, we restrict our attention 
to pairs of encoding and allocation policies $(\phi, \psi)$ under which the Markov chain $\bW(\cdot)$ 
is irreducible.



{{\bf Memory-feedback.} We say that the system has {\bf memory-feedback} if the state of the receiver memory, $M_r(t-1)$, is accessible by the encoding policy in time {slot} $t$ for generating the message $X(t)$ and updating the encoder memory state $M_e(t)$. That is, under this assumption, Eq.~\eqref{eq:encoding} would become: 
\begin{equation}
(X(t), M_e(t)) = \phi(\bQ(t-1), \bA(t-1), M_e(t-1), M_r(t-1), U_e(t-1)), \quad t\in \N. 
\label{eq:encoding_w_FB}
\end{equation}
As alluded to in the Introduction, note that the model already assumes a mode of feedback:  the scheduling decision, $\bD(t)$, can be sent to the queues without obstruction, implying that the backward communication from the receiver to the encoder is lossless. Therefore, 
{under the memory-feedback assumption, in addition to sending $\bD(t)$, the allocation policy also includes the state of its own memory in the backward communication.}}

\subsection{Main Performance Metric: Capacity Factor}\label{ssec:cap_fac}

We define in this subsection the main performance metric of this paper, the capacity factor. Fix a pair of encoding and allocation policies, $\phi$ and $\psi$, respectively. We say the system is \emph{stable} if  $\bW(\cdot)$ is positive recurrent.  Define the \emph{maximum capacity region}, $\Lambda$, to be the set of all vectors strictly dominated by the convex hull of the schedule set $\Pi$: 
\begin{equation}
\Lambda \bydef \pconv(\Pi) = \{\blambda \in \rp^N: \blambda < \conv(\Pi)\}.
\label{eq:capRegDef}
\end{equation}
{Note that because the schedule set $\Pi$ satisfies Assumptions \ref{asmp:monotone} 
and \ref{asmp:non-degeneracy}, it is not difficult to see that {$\cl(\Lambda) = \conv(\Pi)$}.}

%

We now define our main performance metric.

\begin{definition}[Capacity Factor] \label{def:cap_fac}
Fix a channel $\calC$, and $k,v \in \zp$. 
\begin{enumerate}
\item Consider encoding policy $\phi\in \Phi_k$ and allocation policy $\psi\in \Psi_{v}$. Define $\tilde{\Lambda}(\phi, \psi)$ to be the {\em capacity region} under the policy pair $(\phi, \psi)$: 
\begin{equation}\label{eq:stability_region}
\tilde{\Lambda}(\phi, \psi) \bydef \{\blambda\in \rp^N:  \mbox{$(\phi, \psi)$ stabilizes the system under the arrival rate vector $\blambda$} \}. 
\end{equation}
We also define the {\em capacity factor of the channel $\calC$ under the policy pair $(\phi, \psi)$}, denoted by 
$\rho^*(\phi, \psi, \calC)$, 
to be
\begin{equation}
\rho^*(\phi, \psi, \calC) = \sup\{\rho {\geq} 0 : \rho \Lambda \subseteq \tilde{\Lambda}(\phi, \psi)\}.
\label{eq:cap_fac1}
\end{equation}
\item  The {\em $(k, v)$-capacity factor of channel $\calC$}, denoted by $\rho^*_{k,v}(\calC)$, is defined to be
\begin{equation}
\rho^*_{k,v}(\calC)  = \sup\left\{\rho^*(\phi, \psi, \calC) : \phi \in \Phi_k, \psi \in \Psi_{v}\right\}. 
\end{equation}
When the context is clear, sometimes we just write the ``capacity factor'' to mean $\rho^*_{k,v}(\calC)$.
\end{enumerate}
\end{definition}

\noindent {\bf Some elementary properties of capacity factor.} 
{Before we proceed, we state some elementary properties of capacity factor.}
First, for any channel, $\calC$, $\rho_{k,v}^*(\calC)$ must be non-decreasing in both $k$ and $v$, 
because the capacity region can never decrease with more memory. 
Furthermore, since $\rho_{k,v}^*(\calC)$ is upper-bounded by $1$ by definition, 
by the Monotone Convergence Theorem, we have the following: 
\begin{align}
\rho^*_{k, \infty} (\calC) \bydef & \lim_{v \to \infty} \rho^*_{k, v}(\calC), \quad \mbox{ and } \quad \rho^*_{k, \infty} (\calC) \geq \rho^*_{k, v} (\calC) \mbox{ for all $v \in \zp$}; ~\text{and}\\ 
\rho^*_{\infty, v} (\calC) \bydef & \lim_{k \to \infty} \rho^*_{k, v}(\calC),  \quad \mbox{ and } \quad \rho^*_{\infty, v} (\calC) \geq \rho^*_{k, v} (\calC) \mbox{ for all $k \in \zp$};~\text{and}\\ 
\rho^*_{\infty,\infty} (\calC) \bydef &\lim_{k\to \infty }  \lim_{v \to \infty } \rho^*_{k, v}(\calC) = \lim_{v\to \infty }  \lim_{k\to \infty } \rho^*_{k, v}(\calC).
\end{align}
In particular,  the limits in which we take $k$, $v$, or both to $\infty$ are well defined. 

\section{Main Results} 
\label{sec:main_res}
We formally state the main results in this section. We begin with two definitions.
\begin{definition}[Informative Channels] 
\label{def:informC}
A channel $\calC$ is said to be \emph{informative} if the corresponding channel matrix, $C$, admits at least two distinct rows. A channel $\calC$ whose channel matrix has identical rows is called {\em uninformative}.
\end{definition}
{The purpose of Definition \ref{def:informC} is to rule out degenerate channels: simply put, a channel is informative if and only if its output is not independent of the input. The next definition speaks to the other extreme by describing channels that are sufficiently noisy. }

\begin{definition}[$\veps$-Majorizing Channels]
\label{def:eps-Major}
Fix $\veps\in (0,1)$. We say that a channel $\calC$ is \emph{$\veps$-majorizing} if its corresponding channel matrix, $C$, can be written as 
\begin{equation}
C = \veps C^0 + (1-\veps)C^1,
\label{eq:eps-dom-channel}
\end{equation}
where $C^0$ and $C^1$ are two row-stochastic matrices, such that $(a)$ the rows of $C^0$ are identical, 
and $(b)$ every column of $C^1$ has at least one zero entry. 
\end{definition}
Roughly speaking, an $\veps$-majorizing channel can be interpreted as 
having at most $\veps$-portion of the channel being ``completely uninformative.'' 
For technical reasons, we will also assume that the ``uninformative portion,'' $C^0$, 
of an $\veps$-majorizing channel, is everywhere positive:\footnote{Intuitively, since $C^0$ is completely uninformative, whether Assumption \ref{asmp:eps_maj} is satisfied or not 
should have little impact on performance. Assumption \ref{asmp:eps_maj} is used in some of the subsequent proofs 
to ensure that states of the chain $\bW(\cdot)$ are ``easily reachable'' from each other. For more details, 
see Appendix \ref{app:thm:kleps<1}.
}
\begin{asmp}\label{asmp:eps_maj}
Let $\calC = (\calX, \calY, C)$ be an $\veps$-majorizing channel, and let $C^0$ be as in \eqref{eq:eps-dom-channel}. 
Then, for all $x\in \calX$ and $y \in \calY$, $C^0_{x, y} > 0$.
\end{asmp}

The following theorem is the main result of this paper. The same results are summarized in Table \ref{tab:main}, where the rows of the table correspond to Items 1 through 3 in the theorem, respectively.  We will assume that the SPII architecture \emph{has memory-feedback} (See Appendix \ref{app:no_feedback} for a discussion a scenario without memory-feedback.)

\begin{theorem}[Characterization of Capacity Factor] 
\label{thm:main}
Fix the number of queues, $N\in \N$ and a finite schedule set,  $\Pi$, {which satisfies Assumptions \ref{asmp:monotone}, \ref{asmp:non-degeneracy} and \ref{asmp:non-degeneracy2}.} Let $\calC$ be an informative channel (Definition \ref{def:informC}). Suppose the system has memory-feedback. The capacity factor, $\rho^*_{k,v}(\calC)$, satisfies the following:\footnote{In this theorem, the notation $k\geq 0$ should be interpreted as a short-hand for $k$ belonging to the extended {non-negative} integers:  $k\in \zp \cup \{\infty\}$.}
\begin{enumerate}
\item \emph{Memory-less receiver}: when $v=0$, we have that 
\begin{equation}
\rho^*_{k,0}(\calC) = \rho^*_{0,0}(\calC), \quad \forall k \geq 0.
\label{eq:main_l0}
\end{equation}
\item  \emph{Finite-memory receiver}: when $v\in \N$, we have that 
\begin{equation}
\rho^*_{k,v}(\calC) =\rho^*_{K(\Pi, \calX),v}(\calC), \quad \forall  k  \geq K(\Pi, \calX). 
\label{eq:main_lfin}
\end{equation}
where $K(\Pi,\calX) \in \N$ depends only on the structure of the schedule set, $\Pi$, and the channel's input alphabet, $\calX$. 
\item  \emph{Infinite-memory receiver}: when $v=\infty$, we have that
\begin{equation}
\rho^*_{k,\infty}(\calC) = \rho^*_{0,\infty}(\calC)= 1,  \quad \forall  k \geq 0.
\label{eq:main_linf}
\end{equation}
\item Suppose, in addition, that the channel is also $\veps$-majorizing for some $\veps>0$ (Definition \ref{def:eps-Major}), 
{and it satisfies Assumption \ref{asmp:eps_maj}.} Then, for all $v < \infty$, we have
\begin{equation}
\rho_{k,v}^*(\calC) <1, \quad \forall   k \geq 0. 
\label{eq:main_epsmaj}
\end{equation}

\end{enumerate}

\end{theorem}

{\noindent{\bf Remark (The Importance of Memory).} 
At this point, let us provide some remarks regarding the fundamental importance of memory in our model. 
In this paper, we are primarily concerned with the stability analysis of dynamic control policies 
in the SPII. Similar to the stability analysis of conventional Stochastic Processing Networks, 
our problem can be viewed as a relaxed version of an infinite-horizon average-cost Markov Decision Process (MDP): 
instead of trying to minimize the long-run average total queue size, we are concerned with the {\em a priori} 
simpler question of whether the long-run average total queue size can be made finite. 
Since an average-cost MDP typically admits optimal stationary policies that 
do not require additional auxiliary memory, it may seem natural to expect the same of an optimal pair 
of encoding and allocation policies in SPII as well. There is, however, a caveat: while for any fixed instance of SPII (hence a fixed MDP)
this may be true, 
achieving a large capacity region, on the other hand, requires us to identify a single policy pair that performs well across a \emph{set} of different MDPs (parameterized by the arrival rate vector, in our case), which, in general, cannot be accomplished by a single stationary policy (cf.~Definition \ref{def:cap_fac}). 
Indeed, in order for a single policy pair to perform well across a diverse set of problem instances, additional memory is necessary 
for the policy pair to keep track of relevant information and adapt to the specific instance over time. Viewed from this angle, the throughput optimality of the original Max-Weight policy  for SPNs (which is stationary) is a rather remarkable and surprising result \citep{tassiulas1992stability}. However,  the Max-Weight policy crucially relies on being able to fully observe the queue lengths, and, unfortunately, under imperfect information, our results demonstrate that it is no longer possible to achieve maximal capacity region without memory. In particular, Theorem \ref{thm:main} shows that receiver memory is {necessary} for a policy to obtain a large capacity region (Items 3 and 4), while encoder memory seems to be less crucial (Items 1 through 3). A major open problem is that, when the receiver has finite but non-zero memory, whether the encoder memory is needed {at all}, i.e., whether $\rho^*_{0,v}(\calC)=\rho^*_{\infty,v}(\calC)$ for $1\leq v < \infty$.
}

 \section{Related Literature} 
\label{sec:lit_review}
The challenges in obtaining reliable and timely access to state information have long been recognized in large-scale dynamic resource allocation problems. One prominent example  is  the celebrated ``power-of-two-choices'' (PoT) routing algorithm \citep{vvedenskaya1996queueing, mitzenmacher2001power} for load-balancing. Designed to address the lack of full queue-length information in a system with a large number of parallel queues, the PoT algorithm  routes an incoming job to the shorter one between two randomly sampled queues. The same design consideration underlies pull-based variants of PoT \citep{badonnel2008dynamic, lu2011join, stolyar2015pull, stolyar2017pull}, and the partially centralized scheduling policy by \citet{tsitsiklis2012power} that has access to {complete} queue-length information only a small fraction of the time.  Beyond the realm of computer networks, information constraints are also prominent in systems with humans in the loop. For instance, communication failures and misunderstanding between physicians and nurses have been cited as a leading cause of adverse events in healthcare, and specialized messaging and decision protocols have been developed to minimize the impact of errors \citep{manojlovich2015effect}; see \citet{nursehandover, traumahandover} for other examples of information loss among healthcare providers.  While information constraints play a central role in 
{the aforementioned models and applications}, in contrast to our work, they often serve as an implicit motivation behind a chosen design, rather than an explicit  constraint with respect  to which an optimal policy is to be identified. As a result, there has been little understanding as to what policies are ``optimal'' for a given level of information availability, and what the fundamental impact information has on system performance. 

Taking a more principled approach to policy design, several recent papers have aimed at rigorously {quantifying} the performance impact of information in dynamic resource allocation. \citet{gamarnik2018delay} characterize how the average delay in a load-balancing system scales depending on the rate of messaging between the dispatcher and the servers, as well as the size of the dispatcher memory. In the context of queueing admission control, \citet{spencer2014queuing} and \cite{xu2015necessity} quantify how the system's optimal heavy-traffic delay scales as a function of the amount of future information available. In contrast to our approach, however, these papers largely focus on a {specific} model of information constraint, e.g., captured by the rate of messaging or length of the lookahead window, while our framework allows for a substantially more general family of information models, achieved by using different channels. To the best of our knowledge, the present paper is one of the first attempts at rigorously establishing the link between information and the performance of a resource allocation system at this level of generality. 

At a high-level, our framework is {partly} inspired by information theory, and more specifically, the research on feedback control under communication constraints \citep{tatikonda2004control,sahai2006necessity}; see \citet{yuksel2013stochastic} for a survey. This literature studies the  problem of stabilizing (i.e., minimizing the magnitude of the state) a linear dynamical system of the form: $X_{t+1} = AX_t + N_t + U_t$, where $X_t$ is the state, $N_t$ a noise disturbance, $U_t $ the control action, and $A$ a {gain matrix}, and the decision maker has access to the state $X_t$ only through a rate-limited communication channel, similar to the scenario depicted in Figure \ref{fig:sysDiag}. While our framework also admits a feedback loop over a communication channel, the dynamics in our problem differ fundamentally from those in a linear dynamical system, and consequently, so do the design approaches and analysis. The difference stems from the fact that the state process in a linear dynamical system is driven {multiplicatively} by the  gain matrix, $A$, whereas in our system, it is updated in an {additive} manner (see Eq.~\eqref{eq:dynamics}). 
Consequently,  in the control setting, even when all parameters are known (e.g., gain matrix, noise distribution, etc), an informative channel is still {necessary} for stabilization  \citep[e.g.,][]{tatikonda2004control}.  In sharp contrast, {as we discussed earlier in Section \ref{sec:intro},} if all parameters are known in our SPII, it becomes trivial to stabilize the queues without any feedback: the decision maker can simply choose a stationary, randomized allocation policy that {dominate} the arrival rate vector $\blam$.  

{A major theme on the dynamic control of SPNs concerns the setting where 
the decision maker does not have complete information about the underlying system. 
For example, the classical Max-Weight policy \citep{tassiulas1992stability, DaiLin2005, mandelbaum2004scheduling} is oblivious to detailed statistics of the arrival process. 
There is also a literature that addresses the setting where the service rates (or the service time distributions) 
are not known completely, and the decision maker needs to either learn these parameters or develop scheduling policies that do not depend on service rates \citep{StolyarSSY2012, Robust2017, Dimakis2006, Baharian2011, QBandits2016, KAJS2018, WardArmony2013}. 
Let us note that in all this literature, even though the decision maker has partial or no information on {system parameters}, she has full information on {system states}. 
In contrast, our model only assumes (often severely) noisy observations of system states. 
This fundamental difference requires us to take a very different approach in designing policies, 
and to develop new tools for analyzing them.}

Recently, there have been several works that consider the impact of information delay on the performance of processing systems \citep{AL2019, PRW2019}. 
{Information delay can be thought of as a specific form of imperfect information, and arises endogenously from within the system. 
In contrast, information obfuscation in our model arises exogenously. Consequently, our work uses very different analyses from the literature on information delay.}

On the methodological front, our program involves the development of a host of new techniques, largely from first principles, by combining ideas from areas such as information theory, learning and queueing theory.  As a sub-module of one of the policies we propose, we also create a simple yet powerful generalization of the Max-Weight policy, 
in which individual Markov chains are selected dynamically, in a manner  analogous to how schedules are used in a  conventional Max-Weight policy.

\section{Overview of Proof Techniques}
\label{sec:poverview}

The remainder of the paper is devoted to the proof of Theorem \ref{thm:main}, and we provide in this section an overview of the key ideas. 

 \emph{Item 4 of Theorem \ref{thm:main} (Section \ref{sec:epsMaj}) -  Capacity factor is less than one for $\veps$-majorizing channels}. We show that 
 any level of noise in the form of an $\veps$-majorizing channel always reduces the capacity region. The intuition is that the channel noise causes any output symbol to appear with sufficiently positive probability, 
 {independently from the underlying arrival rate vector,} 
 and thus limits the allocation policy's ability to adapt to different arrival rates
 with sufficient precision {(Lemma \ref{lem:communicating} of Appendix \ref{app:thm:kleps<1})}. The proof employs a {lifting} argument, whereby we carry out the analysis in a higher-dimensional product space for the output alphabet, with the added dimension capturing the realization of the channel noise. A coupling argument is then used to show that noise reduces capacity region. 
{The non-degeneracy Assumption \ref{asmp:non-degeneracy2} on the schedule set $\Pi$ 
ensures that there are at least two distinct, 
``extremal'' arrival rate vectors in the maximum capacity region $\Lambda$, 
which, due to the limited adaptability of the allocation policy, cannot be both stabilized, 
implying that the capacity factor is less than one 
(see e.g., proof of Theorem \ref{thm:kleps<1} in Appendix \ref{app:thm:kleps<1} 
and proof of Theorem \ref{thm:epsMajChar} in Appendix \ref{app:thm:epsMajChar}). 
Let us also note that for the special case of a memoryless receiver ($v = 0$), we were able to 
obtain a tight characterization of the corresponding capacity factor under any $\veps$-majorizing channel, 
by providing an achievable upper bound (Theorems \ref{thm:epsMajChar} and \ref{thm:epsMajChar2}).}

 \emph{Items 1 and 2 of Theorem \ref{thm:main} (Sections \ref{sec:memoryless} and \ref{sec:finRecMem}) - Memoryless and Finite-Memory Receivers}. This  is the most technically challenging part of our program. We begin with the simple case of a memoryless receiver ($v=0$), where the size of the encoder memory $k$ has {no effect} on the capacity factor 
 {(Theorem \ref{thm:stationaryCap} of Section \ref{sec:memoryless})}. The key intuition is a {change of perspective}: instead of {focusing on} the receiver (allocation policy) as the one making scheduling decisions based on noisy information, it turns out that the correct way to design the system is to treat the {encoding policy} as the more ``intelligent'' of the two policies that conducts Max-Weight-like scheduling, where the encoding policy treats the set of input symbols, $\calX$, as its ``scheduling actions.'' By formulating a transformed, but equivalent, scheduling problem from the perspective of the encoder, we then use a version of the Max-Weight policy to establish stability. 

For the general case of $v >0$, we obtain a slightly weaker result than that of $v=0$, showing that the {encoder's} memory size becomes irrelevant after a {finite threshold}, $K(\Pi,\calX)$.  The proof builds on the same intuition of viewing the {encoder}, {instead of the receiver}, as the main decision maker. However, a non-trivial receiver memory will mandate a substantially more sophisticated argument. This is because in each time slot, the induced service action
no longer depends solely on the input signal $X(t)$, as in the case of memoryless receiver; it will now also depend on the state of the receiver memory, which is by itself a stochastic process, and hence conventional Lyapunov arguments for Max-Weight cannot be applied. Instead, we will formulate a generalized version of the conventional Max-Weight policy, dubbed the \emph{Episodic Max Weight (EMW)} policy,  where the encoding policies switches between a family of \emph{Markov chains}, as opposed to input symbols, in a manner that is analogous to how schedules are used in conventional Max-Weight {(Section \ref{ssec:EMW})}. The stability proof heavily exploits a certain conditional independence properties among different elements of the overall process $\bW(t)$ {(proof of Proposition \ref{prop:simple} in Appendix \ref{app:prop:simple})}, 
which in turn {was a result of} the feedback structure of the SPII. 

\emph{Item 3 of Theorem \ref{thm:main} (Appendix \ref{app:sec:infRecMem}) - Infinite-Memory Receiver}. The last part of the proof shows that as the receiver memory size $v \to \infty$, the capacity factor always converges to $1$, regardless of the size of encoder memory. The argument is relatively straightforward compared to the other parts. Since the receiver has abundant memory in this regime, the main idea will be shifting the burden of decision-making {back} to the receiver. We construct an \emph{Episodic Greedy Learning (EGL)} policy, where the receiver first estimates the arrival rates from the noisy messages, and subsequently deploys a randomized schedule that {dominates the estimated arrival rate vector in expectation. 
With more memory, the receiver is able to estimate arrival rates more accurately, 
leading to capacity factors that are arbitrarily close to $1$.}

\section{Preliminaries}
\label{sec:prelim}
The main purpose of this section is to establish some results and conventions that will be 
used throughout the remainder of the paper. Section \ref{ssec:abstractMW} introduces a generalized formulation of the Max-Weight policy, whose stability properties will be used as a sub-module in subsequent proofs.  Since our primary focus is on stability, we will often be concerned with 
the question of whether the {long-run average} service rates dominate the arrival rates. 
Section \ref{ssec:station_rate} formalizes the notion of long-run average service rates for our model, 
which {will be} used extensively in later sections.

\subsection{A Generic Max-Weight Stability Theorem}\label{ssec:abstractMW}
In this section, we present a simple generalization of the stability result of the celebrated Maximum Weight (Max-Weight) policy \citep{tassiulas1992stability} to a class of systems that is more general than those typically seen in prior literature  \citep[e.g.,][]{stolyar2004maxweight, DaiLin2005, shah2012switched}.  This result, Proposition \ref{prop:absMW}, will be used as a basic building block in our subsequent proofs. 
We first describe the setup, 
and then present the stability result. 
The proof is a simple modification of the standard stability proof of the Max-Weight policy, 
which we include {in Appendix \ref{app:prop:absMW}} for completeness.


Consider a discrete-time, irreducible Markov chain $\{\bZ(t)\}_{t\in \zp}$ 
with two components $\bQ(\cdot)$ and $G(\cdot)$, 
so that for any time $t$, $\bZ(t) = \left(\bQ(t), G(t)\right)$. 
Here $G(\cdot)$ takes value in a finite set $\calG$, 
and $\bQ(\cdot) \in \zp^N$ evolves according to the following dynamics:  
\begin{equation}\label{eq:dynamics_residual}
\bQ(t)  = (\bQ(t-1)-\bD(t))^+ + \bA(t) - \bR(t),  \quad  t\in \N. 
\end{equation}
Here, $\bA(t)$, $\bD(t)$ and $\bR(t)$ are all random vectors taking values in $\zp^N$. 
It is useful to think of the system dynamics in the following way. 
During each time slot $t$, upon observing the current queue size vector $\bQ(t-1)$, 
the system makes the service allocation decision $\bD(t)$, 
which is used twice in the current slot. 
First, it is used as much as possible to reduce the queue sizes $\bQ(t-1)$, 
as represented by the term $(\bQ(t-1)-\bD(t))^+$. 
There may be residual services left from this first use of $\bD(t)$. 
Then, arrivals $\bA(t)$ take place, 
and some of the residual services may be used to serve the arrivals $\bA(t)$;  { this portion of the residual services is denoted by $\bR(t)$.}

More formally, we require the random vectors $\bA(t)$ to be i.i.d. with finite second moment, 
and independent from the rest of the system, with 
\begin{equation}
\E(\bA(1)) = \balpha = (\alpha_1, \ldots, \alpha_N) \in \rp^N. 
\end{equation}
$\bD(t)$ are also required to have finite second moments, which are uniformly bounded above by a constant 
that does not depend on the time index $t$. Furthermore, given $\bZ(t)$, $\bD(t+1)$ is conditionally independent from the past history $\{\bZ(s): s< t\}$. 
$\bR(t)$ are required to satisfy the following properties w.p.$1$: 
\begin{itemize}
\item[(a)] {$\bR(t) \leq \min\{\bD(t), \, \bA(t)\}$}, and $\bR(t)$ guarantees that $\bQ(t+1) \in \zp^N$; and 
\item[(b)] given $\bZ(t)$, $\bR(t+1)$ is conditionally independent from the past history $\{\bZ(s): s< t\}$.
\end{itemize}
Let us provide some remarks about the differences between the process $\bZ(\cdot)$ 
and those in prior literature, as well as how these differences are used in the present paper.  
Compared to other works on SPNs ~\citep[e.g.,][]{shah2012switched}, 
$\bZ(\cdot)$ has the additional component $G(\cdot)$, 
which is included to model the signals $X(t)$ and the memory contents $M_e(t)$ and $M_r(t)$. 
The ``residual service'' term $\bR(t)$, also absent from related prior literature, 
will be used in Section \ref{ssec:EMWregion} to describe the dynamics of $\bW(\cdot)$ (recall \eqref{eq:W}) 
under a so-called {Episodic Max-Weight policy}. 

We are now ready to state the stability result. The proof is given in Appendix \ref{app:prop:absMW}.
\begin{proposition}
\label{prop:absMW}
Let $\calD$ be a non-empty, finite subset of $\rp^N$ that satisfies {Assumption \ref{asmp:non-degeneracy}}.
Define the sets 
\begin{equation}
\calD^*(t) \bydef \argmax_{\bd \in \calD} \left<\bQ(t), \bd \right>, \quad t\in \zp. 
\end{equation}
Suppose that $\balpha \in \pconv(\calD)$, and for all $t\in \zp$, 
\begin{equation}
\E(\bD(t+1) \bbar \bZ(t)) \geq \bd^*, \quad \mbox{for some $\bd^* \in  \calD^*(t)$ almost surely}. 
\end{equation}
Then $\bZ(\cdot)$ is positive recurrent.   
\end{proposition}
An immediate corollary of Proposition \ref{prop:absMW} is {the following.}
\begin{corollary}\label{cor:absMW}
Suppose that the residual services $\bR(t)$ are zero w.p.1, so that $\bQ(\cdot)$ 
follows the dynamics given by \eqref{eq:dynamics}. Let $\calD$, $\calD^*(\cdot)$ and $\balpha$ 
be exactly as in Proposition \ref{prop:absMW}. Then $\bZ(\cdot)$ is positive recurrent.
\end{corollary}

\subsection{Stationary Service Rates {in SPII}}
\label{ssec:station_rate}
{In conventional SPN systems, it is a simple fact that 
a stabilizing scheduling policy induces stationary service rates that dominate corresponding arrival rates. 
The presence of channel, encoder and receiver in the SPII architecture slightly complicates 
the notion of stationary service rates. {For this reason, we provide in this subsection some discussion on how stationary service rates will be defined.}

Consider a channel $\calC = (\calX, \calY, C)$, constants $k, v \in \zp$, 
and a policy pair $(\phi, \psi)$ with $\phi \in \Phi_k$ and $\psi \in \Psi_{v}$. 
Suppose that when the arrival rate vector is $\blambda$, 
the process $\bW(\cdot)$ is positive recurrent under the policy pair $(\phi, \psi)$. 
Then, $\bW(\cdot)$ has a unique stationary distribution, 
whose corresponding probability we denote by $\pb^{\infty}(\cdot)$.

Next, we define some further notation. 
First, for each $x \in \calX$, define
\begin{equation}\label{eq:stationary_x}
\gamma_x \bydef \pb^{\infty}(X(t) = x)
\end{equation}
to be the stationary probability of sending signal $x$, {and let $\bgamma = (\gamma_x)_{x\in \calX}$}. 
By the ergodicity of the Markov chain $\bW(\cdot)$, we also know that w.p.$1$,
\begin{equation}\label{eq:longrun_x}
\gamma_x = \lim_{T\to \infty} \frac{1}{T} \sum_{t=1}^T \mathbb{I}_{\{X(t)=x\}}, 
\end{equation}
{where $\gamma_x$ represents} the long-run average fraction of time that the sent signal is $x$.
Second, for each $y \in \calY$ and $\bd \in \Pi$, the following conditional probability is well-defined: 
\begin{equation}\label{eq:psi_matrix}
\Theta_{y,\bd} = \pb^{\infty} (\bD(t) = \bd \bbar Y(t) = y). 
\end{equation}
We use $\Theta$ to denote the $c_{\calY} \times |\Pi|$-matrix $(\Theta_{y, \bd})_{y\in \calY, \bd \in \Pi}$, 
and call it the \emph{rate allocation matrix}.
Finally, we also represent the set of allowable schedules in a matrix form. Let $S = (S_{\bd, i})$ be the $|\Pi|\times N$ {\emph{schedule matrix}} where
\begin{equation}\label{eq:schedule_matrix}
S_{\bd, i} = \mbox{ number of services offered to queue $i$ when schedule $\bd$ is chosen.}
\end{equation}
Let us note that in general, $\bgamma$ and $\Theta$ may depend on $\blambda$, 
whereas $S$ does not. 
With the preceding notation in mind, we define the {\em stationary service rate vector $\bmu = \bmu(\phi, \psi, \blambda)$} to be
\begin{equation}\label{eq:stationary_service_rates}
\bmu = \bmu(\phi, \psi, \blambda) = \bgamma C \Theta S, 
\end{equation}
so that for each $i$, $\mu_i$ represents the long-run average service rate offered to queue $i$. 
The following lemma is a simple but useful fact. The proof follows simply from the fact that $\bW(\cdot)$ is positive recurrent.
\begin{lemma}\label{lem:stationary_service_rates}
Let $\calC = (\calX, \calY, C)$, $\blambda$, and $\bW(\cdot)$ be described as above, 
and $\bmu$ be defined by \eqref{eq:stationary_service_rates}. Then, 
$\bmu \geq \blambda$.
\end{lemma}

\section{Capacity Factors Are Non-Trivial For $\veps$-Majorizing Channels}
\label{sec:epsMaj}


In this section, we will establish Item 4 of Theorem \ref{thm:main}, that {the capacity factor as a performance metric is} non-trivial for $\veps$-majorizing channels 
(recall Definition \ref{def:eps-Major}), i.e., that they are in general less than $1$. {Towards this end, we prove the following:}
\begin{theorem}\label{thm:kleps<1}
Consider a system with schedule set $\Pi$ that satisfies Assumptions 
\ref{asmp:monotone}, \ref{asmp:non-degeneracy} and \ref{asmp:non-degeneracy2}. 
Fix any finite $v \in \zp$, $k \geq 0$ $($recall that $k \geq 0$ means $k \in \zp \cup \{\infty\}$$)$, and $\veps \in (0, 1)$. 
Then, for any $\veps$-majorizing channel $\calC=(\calX, \calY, C)$, 
\begin{equation}\label{eq:kleps<1}
\rho^*_{k, v}(\calC) < 1.
\end{equation}
\end{theorem}\
The main idea behind the proof of Theorem \ref{thm:kleps<1} is that because the channel is $\veps$-majorizing, 
each combination of output symbol $Y(t)$ and receiver memory content $M_r(t)$ appears with sufficiently 
positive probability. As a result, no allocation policy (with finite memory) is {adaptive enough} to 
stabilize all arrival rate vectors in $\Lambda$. The proof is given in Appendix \ref{app:thm:kleps<1}.

{The next theorem states a stronger result, for the special case of memoryless receiver, i.e., $v = 0$.
It gives a tight characterization of the $(k, 0)$-capacity factor of any $\veps$-majorizing channel, 
by providing an achievable upper bound, $\rho(\veps, \Pi)$, on the capacity factors.}

\begin{theorem}
\label{thm:epsMajChar}
Consider a system with schedule set $\Pi$ that satisfies  
Assumptions \ref{asmp:monotone}, \ref{asmp:non-degeneracy} and \ref{asmp:non-degeneracy2}. 
Then, for any $\veps\in (0, 1)$, there exists a constant $\rho(\veps, \Pi) < 1$, 
{which only depends on $\veps$ and $\Pi$}, such that: 
\begin{enumerate}
\item For any $\veps$-majorizing channel $\calC=(\calX,\calY,C)$, and for any $k \geq 0$, 
\begin{equation}
\rho^*_{k,0}(\calC) \leq \rho(\veps, \Pi). 
\end{equation}
\item Conversely, there exists an $\veps$-majorizing channel, $\calC$, such that 
\begin{equation}
\rho^*_{0,0}(\calC) = \rho(\veps, \Pi). 
\end{equation}
\end{enumerate}
\end{theorem}

{Next, we state Theorem \ref{thm:epsMajChar2}, a special case of Theorem \ref{thm:epsMajChar}, 
for systems of parallel queues with a single server, where we derive a simple, 
explicit expression for the bound $\rho(\veps, \Pi)$. 
The proof of Theorem \ref{thm:epsMajChar} is more abstract than that of Theorem \ref{thm:epsMajChar2} 
and involves using general properties of convex {polyhedra}, 
but it follows a similar argument and is provided in Appendix \ref{app:thm:epsMajChar}.} 

While more restricted in scope, Theorem \ref{thm:epsMajChar2} best illustrates the key intuition present in the more general version of the result. 
Consider the following system that consists of one server and $N$ parallel queues, where the queues are indexed by $[N]$. In each time slot, the server picks one queue, so that exactly one job departs from the chosen queue if it is non-empty, and no departure occurs in the system, otherwise. Thus, the schedule set $\Pi$ is given by
\begin{equation}
\Pi = \{\bZero\} \cup \{\be^{(i)}\}_{i=1}^N,
\end{equation}
where $\bZero$ is the zero vector, and $\be^{(i)}$ is the standard $i$th unit vector, $i \in [N]$. 
It is easy to see that the capacity region of this system is given by 
\begin{equation}
\Lambda = \lt\{\blambda \geq 0: \sum_{i=1}^N\lambda_i < 1 \rt\}. 
\end{equation}



%
The following theorem provides a tight characterization on the capacity factors of $\veps$-majorizing channels, 
when the allocation policy is memoryless.
\begin{theorem}
\label{thm:epsMajChar2}
Fix $\veps\in (0,1)$. Consider a single-server system with $N$ parallel queues. Then, the following hold.
\begin{enumerate}
\item For any $\veps$-majorizing channel, $\calC=(\calX,\calY,C)$, and any $k \geq 0$, 
\begin{equation}
\rho^*_{k,0}(\calC) \leq 1-\veps(1-N^{-1}). 
\end{equation}
\item Conversely, there exists an $\veps$-majorizing channel, $\calC$, such that 
\begin{equation}
\rho^*_{0,0}(\calC) = 1-\veps(1-N^{-1}). 
\end{equation}
\end{enumerate}
\end{theorem}

The proof of Theorem \ref{thm:epsMajChar2} is given in Appendix \ref{app:thm:epsMajChar2}.

\section{Memoryless Receiver}
\label{sec:memoryless}
In this section, we show {Item 1 of Theorem \ref{thm:main}:} if the allocation policy is memoryless (i.e., $v=0$), then there exists some memoryless encoding policy $\phi\in \Phi_0$, which achieves the optimal capacity factor among all encoding policies, memoryless or not. In short, {when the receiver does not have memory, then one will not benefit from any memory at the encoder, either.}

\begin{theorem}\label{thm:stationaryCap} 
Fix any channel $\calC=(\calX,\calY, C)$. We have that
\begin{equation}
\rho^*_{0,0}(\calC) = \rho^*_{\infty,0}(\calC). 
\end{equation}
\end{theorem}

{The proof of Theorem \ref{thm:stationaryCap} advances a design philosophy that will form the core of the proofs in subsequent sections. The key is to view the \emph{encoding policy},  rather than the allocation policy, as the true ``decision maker,'' even though the latter physically selects the schedule. Concretely, note that for any {fixed} allocation policy, the expected allocation is solely determined by the input symbol. As such, we may implement a {Max-Weight-like} policy at the encoder, where the ``maximization'' is performed across the input symbols, as opposed to physical schedules. }

\bpf 
It is clear that {more  encoder memory} cannot degrade performance, i.e., $\rho^*_{0,0}(\calC) \leq \rho^*_{\infty,0}(\calC)$, 
so it suffices to show that $\rho^*_{0,0}(\calC) \geq \rho^*_{\infty,0}(\calC)$. 
For the rest of the proof, we write $\rho$ for $\rho^*_{\infty, 0}(\calC)$ to simplify notation.

{Fix} $\veps>0$. We will show that $\rho^*_{0,0}(\calC) \geq \rho - \veps$ by constructing 
a policy pair $(\phi, \psi)$ with $\phi \in \Phi_0$ and $\psi \in \Psi_0$, under which 
$(\rho-\veps) \Lambda \subseteq \tilde{\Lambda}(\phi, \psi)$. 
By the definition of $\rho^*_{\infty,0}(\calC)$, 
there exist $k \in \N$, encoding policy $\phi_k \in \Phi_k$ and allocation policy $\psi \in \Psi_0$, 
such that $\rho^*(\phi_k, \psi, \calC)$, the capacity factor of $\calC$ under $(\phi_k, \psi)$, satisfies $\rho^*(\phi_k, \psi, \calC) \geq \rho - \veps/3$. {We will fix such an allocation policy, $\psi$, in the remainder of the proof.}

Let $\blambda \in (\rho-2\veps/3)\Lambda \subseteq \tilde{\Lambda}(\phi_k, \psi)$, 
and consider a system with arrival rate vector $\blambda$, 
operating under the policy pair $(\phi_k, \psi)$. 
Let the stationary service rate vector $\bmu = \bgamma C \Theta S$ be defined by \eqref{eq:stationary_service_rates}, 
and $\bgamma$, $\Theta$ and $S$ defined by \eqref{eq:stationary_x}, \eqref{eq:psi_matrix} and \eqref{eq:schedule_matrix} respectively. 
Then, by Lemma \ref{lem:stationary_service_rates}, $\bmu\geq \blambda$. 
Furthermore, because the allocation policy $\psi$ is memoryless, 
the matrix $\Theta$ depends only on the policy $\psi$, but not on $\blambda$.
Thus, the matrix $C \Theta S$ is fixed, and
$\blambda \leq \conv\big(\what{\calS}\big)$, where $\what{\calS}$ 
denotes the set consisting of the rows of $C \Theta S$: 
\begin{equation}
\what{\calS} = \{(C \Theta S)_x: x\in \calX\}.
\end{equation}
Here $(C \Theta S)_x$ represents the $x$th row of $C \Theta S$. 
Since $\blambda$ is arbitrary except $\blambda \in (\rho-2\veps/3)\Lambda$, 
this implies that 
\begin{equation}\label{eq:cap_region00}
(\rho-\veps)\Lambda \subseteq \pconv\big(\what{\calS}\big).
\end{equation}

We are now ready to define the memoryless encoding policy $\phi \in \Phi_0$ 
and prove that $\rho^*(\phi, \psi, \calC) \geq \rho - \veps$.
For all time $t \in \N$, the sent signal $X(t)$ is a deterministic function of $\bQ(t-1)$, 
defined to be 
\begin{equation}\label{eq:MW00}
X(t) = \phi(\bQ(t-1)) \in \argmax_{x \in \calX} \left<\bQ(t-1), \E\lt(\bD(t) \bbar X(t) = x\rt) \right>,
\end{equation}
where 
\begin{equation}
\E\lt(\bD(t) \bbar X(t) = x\rt) = (C \Theta S)_x, 
\end{equation}
with  ties are broken arbitrarily. Note that $(C \Theta S)_x$ is the vector of expected 
allocation, conditional on alphabet $x$ being chosen to be the signal. {As such \eqref{eq:MW00} can be viewed as a Max-Weight policy implemented by the encoder, where the ``schedules'' indeed correspond to the set of input symbols.}

Let us summarize the encoding-allocation policy pair $(\phi, \psi)$ that we have constructed so far: 
\begin{enumerate}[(1)]
\item encoding policy $\phi$: send a signal $x^*$ according to Eq. \eqref{eq:MW00}. Note that the optimization 
in Eq.~\eqref{eq:MW00} only involves the current queue lengths, so $\phi$ is memoryless. 
\item allocation policy $\psi$: one that corresponds to the allocation policy matrix $\Theta$ 
defined by \eqref{eq:psi_matrix}. 
\end{enumerate}
Under the pair $(\phi, \psi)$, the expected services offered at time $t$ is given by $(C \Theta S)_{X(t)}$. 
The set $\what{\calS}$ satisfies Assumption \ref{asmp:non-degeneracy} because of Eq.~\eqref{eq:cap_region00}, 
and the fact that $\Lambda$ satisfies Assumption \ref{asmp:non-degeneracy}.
Thus, we can apply Corollary \ref{cor:absMW} to the process $\bW(\cdot)$, 
and conclude that for any $\blambda \in (\rho-\veps)\Lambda \subseteq  \pconv\big(\what{\calS}\big)$, 
$\bW(\cdot)$ is positive recurrent under the pair $(\phi, \psi)$. 
This implies that $\rho^*_{0,0}(\calC) \geq \rho^*(\phi, \psi, \calC) \geq \rho - \veps$. 

Since $\veps>0$ is arbitrary, we have that $\rho^*_{0, 0}(\calC) \geq \rho = \rho^*_{\infty,0}(\calC)$. 
This concludes the proof. 
\qed

\section{Finite-Memory Receiver}
\label{sec:finRecMem}

{We now study the case where the receiver memory size, $v$, is finite, and prove Item 2 of Theorem \ref{thm:main}, 
{that there exists a fixed threshold $K(\Pi, \calX)$ on the memory size of the encoder, 
beyond which any additional encoder memory does not improve the capacity factor of the system, for any fixed receiver memory size.} 
We will build on the key insight from Section \ref{sec:memoryless}, that is, to view the encoding policy as the primary decision maker, who employs a Max-Weight-like policy by choosing among the  input signals  (Eq.~\eqref{eq:MW00}). The presence of the receiver memory, however, poses a new challenge {for the proof}: The expected allocation is no longer solely a function of the input signal, $X(t)$, but must also depend on the state of the receiver memory, $M_r(t-1)$. Conversely, the input signal $X(t)$ not only affects the allocation vector $\bD(t)$ in the present time slot, but also influences the state of the receiver memory $M_r(t)$, and through such influence exerts a long-lasting impact on future allocations. 
{Consequently,} we cannot hope to simply imitate the policy used for {the case} when $v=0$ (i.e., Eq.~\eqref{eq:MW00}) and choose input symbols in a myopic manner: Any successful policy must also take into account the receiver memory's dynamics in the long run.}

{With these challenges in mind, we will introduce an important instrument in this section: \emph{simple encoding policies}. These are a family of encoding policies that do not take the queue lengths as input (hence ``simple''). Just as we took a step of abstraction in Section \ref{sec:memoryless} by creating a Max-Weight policy that treats the set of channel input symbols as its available actions, we will now abstract even further to design an (episodic) Max-Weight policy that views a set of simple encoding policies as its available actions, and switches {among} them on a periodic basis. Crucially, when choosing which simple policy to deploy, we will {consider} the corresponding stationary service rates (Section \ref{ssec:station_rate}), and it is the use of the stationary (as opposed to instantaneous) service rates that allows us to successfully induce  the optimal long-term dynamics in the receiver memory. 
{Finally, under the episodic Max-Weight policy, the encoder's memory content is a combination of 
(a) the memory required for the encoder to keep track of the current simple policy it employs, 
and (b) the memory content required of this current simple policy; 
together, they lead to the characterization of the threshold $K(\Pi,\calX)$.}}


The remainder of the section is organized as follows: we first define the notion of {simple encoding policies} in Section \ref{ssec:proj_simple}, 
and establish a key property of simple policies in Proposition \ref{prop:simple}. 
Subsequently, the {\em Episodic Max-Weight} policy, defined in Section \ref{ssec:EMW}, 
makes use of simple policies, 
and is itself used in Section \ref{ssec:EMWregion} to prove Theorem \ref{thm:EMWregion} 
(Item 2 of Theorem \ref{thm:main}). 
Along the way, we also provide an explicit expression for $K(\Pi,\calX)$, 
the threshold beyond which $\rho^*_{k,v}$ are all the same.} Throughout {this section}, we will assume the existence of {memory-feedback}, so that the encoder can 
fully observe $M_r(t)$, the receiver memory, and use it as input to the policy design: 
\begin{equation}
(X(t),M_e(t)) = \phi(\bQ(t-1), \bA(t-1), M_e(t-1), M_r(t-1), U_e(t-1)),  \quad t\in \N.
\end{equation}
\subsection{Projection to Simple Policies}\label{ssec:proj_simple}

\begin{definition}[Simple Encoding Policies] 
\label{def:simpPol}
Fix $v \in \N$ and $\psi \in \Psi_{v}$. An encoding policy $\phi^0$ is called \emph{simple} 
for the allocation policy $\psi$, if 
\begin{enumerate}
\item $\phi^0 \in \Phi_{\lceil \log|\calX| \rceil}$, {and in each time slot $t$, the encoder memory $M_e(t)$, 
is set to equal the current signal $X(t)$, i.e., $M_e(t) = X(t)$}. 
\item $\phi^0$ takes as inputs only $M_e(t-1)$, $M_r(t-1)$ and $U_e(t-1)$, and does not rely on $\bQ(t-1)$ 
{or $\bA(t-1)$}. 
\item $\{(X(t), M_r(t))\}_{t\in \N}$ is a homogeneous irreducible Markov chain under the pair $(\phi^0, \psi)$.
\end{enumerate}
We use $\Phi^S(\psi)$ to denote the set of all simple encoding policies for $\psi$. 
When the context is clear, we sometimes write $\Phi^S$ for $\Phi^S(\psi)$, 
and call a policy simple if it is simple for $\psi$.
\end{definition}
{Let us note that in item $1$ of Definition \ref{def:simpPol}, the requirement that the encoder memory 
be set to the transmitted signal in each time slot may appear to be restrictive. However, as it will turn out 
(Proposition \ref{prop:simple} and Lemma \ref{lem:epsCapkl}), 
we can essentially ``cover'' the capacity region associated with any pair of encoding and allocation policies 
using a sparse set of simple policies thus defined.}

{Next, recall} $\bmu(\phi, \psi, \blambda)$, the stationary service rate vector defined by \eqref{eq:stationary_service_rates}
when the arrival rate vector is $\blambda$, and the system is operating under the policy pair $(\phi, \psi)$. 
If the service rates do not change as a function of $\blambda$, then we write $\bmu(\phi, \psi)$ and omit the dependence on $\blambda$. 
Let us note that for any simple policy $\phi^0$, even though $\bW(\cdot)$ may not be positive recurrent 
under the policy pair $(\phi^0, \psi)$ and arrival rate vector $\blambda$, 
the stationary service rate vector $\bmu(\phi^0, \psi, \blambda)$ 
is still well-defined according to \eqref{eq:stationary_service_rates}, 
because $\{(X(t), M_r(t))\}_{t\in \N}$ is irreducible with a finite state space. 
Furthermore, because $\phi^0$ does not rely on queue lengths when generating signals, 
$\bmu(\phi^0, \psi, \blambda)$ does not depend on $\blambda$, 
in which case we can just write $\bmu(\phi^0, \psi)$. 

The following proposition is the main result of this subsection.

\begin{proposition} 
\label{prop:simple}
Fix $k$, $v \in \N$, $\phi \in \Phi_k$,  $\psi \in \Psi_{v}$, and  $\blambda\in \tilde{\Lambda}(\phi, \psi)$. There exists a simple encoding policy $\phi^0 \in \Phi^S$, such that
\begin{equation}
\bmu(\phi^0, \psi) \geq \blambda. 
\end{equation}
\end{proposition}
{Proposition \ref{prop:simple} provides the technical foundation for the design of the EMW policy 
in Section \ref{ssec:EMW}. 
With the proposition on hand, we can ``approximately'' cover the entire schedule set $\Pi$ 
(see Lemma \ref{lem:epsCapkl} for details), and the EMW policy will switch between simple policies in this covering, 
similar to how schedules are chosen in conventional Max-Weight. 
The proof of Proposition \ref{prop:simple} heavily exploits the conditional independence 
of $X(t)$ and $M_r(t)$ given $X(t-1)$ and $M_r(t-1)$, for any $t$; see Appendix \ref{app:prop:simple} for the detailed proof.}

\subsection{Episodic Max Weight}\label{ssec:EMW}
We present in this section our main policy, Episodic Max Weight, for the regime of $0<v<\infty$, which will employ the simple policies (Definition \ref{def:simpPol}) as basic building blocks.  Recall from Eq.~\eqref{eq:capRegDef} that $\Lambda = \pconv(\Pi)$, 
the set of points strictly dominated by the convex hull of $\Pi$.
Let $\calF$ be the set of all maximal elements of $\Pi$, which we denote by 
$\calF = \lt\{\bmu^\calF_1, \ldots, \bmu^\calF_{|\calF|}\rt\}$. 
Then, it is easy to see that we also have $\Lambda = \pconv(\calF)$. The next lemma is a useful structural result concerning the simple policies; the proof is given in Appendix \ref{app:lem:epsCapkl}. 
\begin{lemma}\label{lem:epsCapkl}
Fix $k, v \in \N$ and $\veps \in (0, \rho^*_{k, v})$. There exist an allocation policy $\psi^\veps \in \Psi_{v}$, 
and a set $\Phi^\veps$ of $|\calF|$ simple encoding policies, 
\begin{equation}
\Phi^\veps\bydef \{\phi_1^\veps, \ldots, \phi_{|\calF|}^\veps\} \subseteq \Phi^S, 
\end{equation}
such that for each $i \in [|\calF|]$, the chain $\left(X(\cdot), M_r(\cdot)\right)$ is irreducible and aperiodic 
under $(\phi^\veps_i, \psi^\veps)$, and 
\begin{equation}
(\rho^*_{k,v}-\veps) \Lambda \subseteq \pconv\lt(\bmu(\Phi^\veps, \psi^\veps) \rt). 
\end{equation}
\end{lemma}

{\bf The Episodic Max Weight Policy.}  Fix $k, v \in \N$ and $\veps \in (0, \rho^*_{k, v})$. Let $\Phi^\veps$ and $\psi^\veps \in \Psi_{v}$ be defined as in Lemma \ref{lem:epsCapkl}. 
The Episodic Max Weight (EMW) encoding policy  is parameterized by an expected episode length parameter, $B \in \N$, whose memory requirement is 
\begin{equation}
K(\Pi, \calX) =\log(|\calF|) +  \log(|\calX|). 
\end{equation}
In every time slot, the EMW policy applies a simple encoding policy from the set $\Phi^\veps$ to generate the signal, 
where $\Phi^\veps$ is defined as in Lemma \ref{lem:epsCapkl}.
The encoder memory stores only two types of information: 
\begin{enumerate}
\item The index of the simple encoding policy in $\Phi^\veps$ that is currently used to generate the signal $X(t+1)$. 
Since $\Phi^\veps$ contains $|\calF|$ simple policies, this can be implemented using $\log(|\calF|) $ bits of memory. 
\item The current signal $X(t)$. This can be implemented using $\log(|\calX|)$ bits of memory. 
\end{enumerate}
We denote the first $\log(|\calF|)$ bits and last $\log(|\calX|)$ bits of $M_e(t)$ by $M^\calF_e(t)$ and $M^\calX_e(t)$, respectively.

\begin{definition}[$B$-Episodic Max Weight ($B$-EMW)]\label{def:BEMW}
Fix $B  \in \N$. Let $\{Z^p_t\}_{t\in \N}$ be a sequence of i.i.d.~Bernoulli random variables with $\pb(Z^P_1 = 1) = \frac{1}{B}$, 
independently from everything else. 
For each $t\in \N$: 
\begin{enumerate}
\item  If $Z^P_t=1$, let
\begin{equation}
i^* \in \argmax_{i =1, \ldots, |\calF|} \lt< \bQ(t), \bmu(\phi^\veps_i, \psi^\veps)\rt>, 
\end{equation}
with ties broken arbitrarily. Set $M^\calF_e(t) = i^*$.
\item If $Z^P_t=0$, set $M^\calF_e(t)=M^\calF_e(t-1)$. 
\end{enumerate}
Apply the simple encoding policy $\phi^\veps_{M^\calF_e(t)}$ to generate $X(t)$. 
Set $M^\calX_e(t) = X(t)$.
\end{definition}

\subsection{Asymptotic Feasible Region under EMW}\label{ssec:EMWregion}

We now present the main result of this section, which will imply Item 3 of Theorem \ref{thm:main}. 
\begin{theorem}\label{thm:EMWregion}
Let $k,v\in \N$, and $\veps \in (0, \rho^*_{k, v})$. Let $\phi^\veps$ be defined as in Lemma \ref{lem:epsCapkl}, 
and denote by $\phi^B$ the $B$-EMW policy defined in Definition \ref{def:BEMW}. Then, there exists $B^*>0$ such that
\begin{equation}\label{eq:EMWregion}
(\rho^*_{k,v}-2\veps) \Lambda \subseteq \tilde{\Lambda}(\phi^{B}, \psi^\veps), \quad \mbox{for all } B>B^*.
\end{equation}
Because $\phi^B \in \Phi_{K(\Pi, \calX)}$ and Eq. \eqref{eq:EMWregion} holds for all $k$ and $\veps$, we have that 
$\rho^*_{k, v} \leq \rho^*_{K(\Pi, \calX), v}$ for all $k$, so
\begin{equation}
\rho^*_{K(\Pi, \calX), v} (\calC) = \rho^*_{\infty, v}(\calC) . 
\end{equation}
\end{theorem}

\bpf
Denote by $T_n$ the $n$th time that $Z^P_t = 1$, and call the time slots between $T_n$ 
and $T_{n+1}-1$ (including both $T_n$ and $T_{n+1}-1$), the $n$th {\em episode}. 
It is easy to see that the episode lengths are i.i.d geometric random variables with mean $B$, 
and that within each episode, the same simple policy is employed by $\phi^B$. 
Define
\begin{equation}
\bA[n] = \sum_{t =T_{n-1}}^{T_{n}-1}  \bA(t), \quad \bD[n] = \sum_{t =T_{n-1}}^{T_{n}-1} \bD(t), \quad \bQ[n] = \bQ(T_n), \quad \text{and} \quad \bW[n] = \bW(T_n). 
\end{equation}
Because the $Z^P_t$'s are i.i.d Bernoulli random variables which are independent from everything else, 
it is easy to see that $\bW[\cdot]$ is a Markov chain. Furthermore, $\bW[\cdot]$ is of the form 
$\bW[\cdot] = (\bQ[\cdot], G[\cdot])$, where $G[n] = (M_e(T_n), X(T_n), M_r(T_n))$ taking values in a finite-state space, 
and $\bQ[\cdot]$ evolves according to the dynamics
\begin{equation}\label{eq:EMWdynamics}
\bQ[n] = (\bQ[n-1] - \bD[n])^+ + \bA[n] - \bR[n],
\end{equation}
with $\bR[n]$ satisfying properties (a) and (b) in Section \ref{ssec:abstractMW}. 
To verify Eq. \eqref{eq:EMWdynamics}, 
we can think of the allocations $\bD[n]$ as first being used to serve $\bQ[n-1]$, 
the queue lengths at the beginning of the episode, and any residual allocations as being 
used to serve subsequent arrivals in the episode. 
These residual allocations are captured by the term $\bR[n]$.

Denote by $\phi[n]$ the simple policy used in the $n$th episode. 
Recall the set $\Phi^\veps = \{\phi^{\veps}_i : i \in [|\calF|]\}$ of simple policies defined in Lemma \ref{lem:epsCapkl}, 
and that for each $i \in [|\calF|]$, $(X(\cdot), M_r(\cdot))$ is irreducible and aperiodic under $(\phi^\veps_i, \psi^\veps)$.
Since $\Phi^\veps$ is a finite set,  the convergence speeds to the steady-state distribution of $(X(\cdot), M_r(\cdot))$ 
under different $(\phi^\veps_i, \psi^\veps)$, $i \in [|\calF|]$, can be uniformly upper bounded. 
Also recall that the episode lengths are i.i.d geometric random variables with mean $B$, 
which are independent from everything else. 
Thus, there exists $B^*_{\veps}>0$, such that if $B>B^*_{\veps}$, then
\begin{equation}
\E(\bD[n+1] \bbar \bW[n]) \geq B(1-\veps)\bmu(\phi[n], \psi^\veps), \quad \forall n \in \N, \mbox{ almost surely.}
\label{eq:ED-EMW}
\end{equation}
Fix $B>B^*_\veps$, and let $\blambda \in (1-\veps)\lt(\rho^*_{k,v}-\veps\rt) \Lambda$. 
By Lemma \ref{lem:epsCapkl}, 
\begin{equation}
(\rho^*_{k,v}-\veps) \Lambda \subseteq \pconv\lt(\bmu(\Phi^\veps, \psi^\veps)\rt),  
\end{equation}
so 
\begin{equation}
B\blambda \in \pconv\lt(B(1-\veps)\bmu(\Phi^\veps, \psi^\veps)\rt).
\label{eq:Bblambda}
\end{equation}
Also note that
\begin{equation}
\E(\bA[n]) = B\blambda, \quad \forall n\in \N. 
\label{eq:EA-EMW}
\end{equation}
By Eqs. \eqref{eq:ED-EMW}, \eqref{eq:Bblambda} and \eqref{eq:EA-EMW}, and Proposition \ref{prop:absMW}, 
the sampled chain $\bW[\cdot]$ is positive recurrent under $B$-EMW whenever 
\begin{equation}\label{eq:EMWregion1}
\blambda \in (1-\veps)(\rho^*_{k,v}-\veps) \Lambda.
\end{equation}
By \eqref{eq:EMWregion1} and the fact that $(1-\veps)(\rho^*_{k,v}-\veps) \geq \rho^*_{k,v}-2\veps$, 
we establish \eqref{eq:EMWregion} and, hence the theorem as well. 
\qed


\section{Conclusion and Future Work}
\label{sec:conclude}

We proposed in this paper the \emph{Stochastic Processing under Imperfect Information (SPII)} framework to study the impact of information constraints and memories on the capacity of general Stochastic Processing Networks. Using a novel metric, capacity factor, our main theorem characterizes the reduction in capacity region (under ``optimal'' policies) under {all non-degenerate channels}, and across {almost all memory sizes}. Our results lead to a number of architectural insights, including: $(1)$ the presence of a noisy channel {always reduces} capacity, $(2)$ more memories for the allocation policy {always improve} capacity,  and $(3)$ more memories for the encoding policy have {little to no effect} on capacity.

A few interesting questions remain open. Firstly, our  results  leave open the question of whether the encoder memory  has {any} impact on the capacity factor when the receiver memory is finite but {non-zero} ($v >0 $), i.e., whether we can reduce the threshold $K(\Pi,\calX)$ down to $0$. Resolving this question in the {negative} would imply an intricate role of encoder memory: it is useful only if the receiver has {some} memory, but not {too much} or {too little} memory. A {positive} answer, on the other hand, would establish that {only} the receiver memory carries any performance benefits in the SPII systems, which would be a rather strong, and arguably surprising, conclusion.  Secondly, our results currently rely on the assumption of memory-feedback for the regime of finite receiver memory $v$, which we hope can be removed in the future. However, we tend to believe that without memory-feedback, the encoder memory should play a crucial role: it can be used for {estimating} the state of the receiver memory, which, without memory-feedback, is no longer directly observable by the encoding policy. As such, we expect that Item 2 in Theorem \ref{thm:main} will no longer hold as is.  Finally, we believe the SPII framework put forth in this paper can be  extended to studying the impact of imperfect information and memory on other, more refined metrics, such as average delay, and more broadly, to other sequential decision-making problems. 

{\section*{Acknowledgements.} Financial supports from Stanford Graduate School of Business and Chicago Booth School of Business are gratefully acknowledged. 
The authors would also like to thank the anonymous reviewers, whose feedback greatly improved various aspects of the paper.}

\bibliographystyle{apa}
\bibliography{IMS.bib}

\newpage 

 \setcounter{page}{1}
 
\ifx \useplain\undefined
\begin{APPENDICES}
\else
\appendix
\fi

\normalsize

\begin{center}
\Large Supplemental Material for ``Information and Memory in Dynamic Resource Allocation''
\\
\vspace{10pt}
\normalsize Kuang Xu and Yuan Zhong\\
 August, 2019
\end{center}

\section{Proofs}

\subsection{Proof of Proposition \ref{prop:absMW}}
\label{app:prop:absMW}
\bpf
Consider the quadratic Lyapunov function $L: \zp^N \times \calG \to \rp$ defined by
$L(\bZ) = \|\bQ\|^2$. 
We are interested in the conditional drift $\E[L(\bZ(t+1)) - L(\bZ(t)) \bbar \bZ(t)]$. We have
\begin{eqnarray}
&& L(\bZ(t+1)) - L(\bZ(t)) \nln
&=&
\|\bQ(t+1)\|^2 - \|\bQ(t)\|^2 \nonumber \\
&=& \sum_{i=1}^N \left\{ \left[ (Q_i(t)-D_i(t+1))^+ + A_i(t+1) - R_i(t+1)\right]^2 - Q_i^2(t)\right\} \nonumber \\
& \leq & \sum_{i=1}^N \left\{ \left[ (Q_i(t)-D_i(t+1))^+ + A_i(t+1)\right]^2 - Q_i^2(t)\right\} 
\label{eq:absMW1}\\
& = & \sum_{i=1}^N \left\{ \left[ (Q_i(t)-D_i(t+1))^+ \right]^2 + A_i^2(t+1) + 2A_i(t+1) (Q_i(t)-D_i(t+1))^+ - Q_i^2(t)\right\} \nonumber \\
& \leq & \sum_{i=1}^N \left\{ \left(Q_i(t)-D_i(t+1)\right)^2 + A_i^2(t+1) + 2A_i(t+1) Q_i(t) - Q_i^2(t)\right\} 
\label{eq:absMW2} \\
&=& \sum_{i=1}^N \left\{ -2Q_i(t)D_i(t+1) + 2A_i(t+1) Q_i(t) + D_i^2(t+1) + A_i^2(t+1)\right\}. 
\label{eq:absMW3}
\end{eqnarray}
Inequality \eqref{eq:absMW1} follows from properties (a) and (b) of the ``residual service'' $\bR(\cdot)$, 
and Inequality \eqref{eq:absMW2} follows from the facts that $\big[(x)^+\big]^2 \leq x^2$ for any $x \in \R$, 
and that for any $x, y \in \rp$, $0 \leq (x-y)^+ \leq x$.

Since both $\bA(t)$ and $\bD(t)$ have second moments that are uniformly upper bounded, 
let us suppose that for some constant $K > 0$, for all time $t$, 
\[
\sum_{i=1}^N \E\left[A_i^2(t) + D_i^2(t)\right] \leq K. 
\]
Conditioning on $\bZ(t)$ and taking expectations on both sides of \eqref{eq:absMW3} gives
\begin{eqnarray*}
\E\left[L(\bZ(t+1)) - L(\bZ(t)) \bbar \bZ(t)\right]
& \leq & -2\left<\E\left[\bD(t+1)\bbar \bZ(t)\right], \bQ(t)\right> + 2\left<\balpha, \bQ(t)\right> + K \\
& \leq & -2\left<\bd^*, \bQ(t)\right> + 2\left<\balpha, \bQ(t)\right> + K,
\end{eqnarray*}
for some $\bd^* \in \calD^*(t+1)$. Since $\balpha \in \rel\left(\pconv(\calD)\right)$, 
there exist constants $\delta \in (0, 1)$, and $p_{\bd} \geq 0$ for each $\bd \in \calD$, 
such that $\sum_{\bd \in \calD} p_{\bd} \leq 1 - \delta$ and $\balpha \leq \sum_{\bd \in \calD} p_{\bd} \bd$. 
Therefore, 
\[
\left<\balpha, \bQ(t)\right> \leq \left<\sum_{\bd \in \calD} p_{\bd} \bd, \bQ(t)\right> 
= \sum_{\bd \in \calD} p_{\bd} \left<\bd, \bQ(t)\right> \leq (1-\delta) \left<\bd^*, \bQ(t)\right>. 
\]
{Thus, 
\[
\E\left[L(\bZ(t+1)) - L(\bZ(t)) \bbar \bZ(t)\right] \leq -2\delta \left<\bd^*, \bQ(t)\right> + K 
\leq -2\delta c \|\bQ(t)\|_{\infty} + K,
\]
for some $c > 0$, 
where the last inequality follows from Assumption \ref{asmp:non-degeneracy}.}

Consider the finite set $\tilde{\calG} \bydef \left\{\bq \in \rp^N: \|\bq\|_{\infty} \leq \frac{K}{2\delta c}+1\right\}\times \calG$. 
Then we have
\[
\E\left[L(\bZ(t+1)) - L(\bZ(t)) \bbar \bZ(t)\right] \leq 
\left\{\begin{array}{ll}
K, & \quad \text{if}\ \bZ(t) \in \tilde{\calG}, \\
-2\delta c, & \quad \text{if}\ \bZ(t) \notin \tilde{\calG}.
\end{array}\right.
\]
The positive recurrence of $\bZ(\cdot)$ then follows from a standard application of the Foster-Lyapunov criteria \citep[e.g.,][]{tassiulas1992stability, hajek2015random}. \qed

\subsection{Proof of Theorem \ref{thm:kleps<1} }
\label{app:thm:kleps<1}

We prove Theorem \ref{thm:kleps<1} in this subsection, and begin with the following technical result on the geometry of the set $\conv(\Pi)$.
\begin{lemma}\label{lem:max_extreme}
Consider a schedule set $\Pi$ that satisfies Assumption \ref{asmp:monotone}, 
and let $\calE$ be the set of extreme points of $\conv(\Pi)$. 
Let $\bd^{(0)}\in \Pi$ be an extreme point that is maximal in $\calE$. Then, $\bd^{(0)}$ is also maximal in $\conv(\Pi)$.
\end{lemma}
\bpf
{Suppose that $\bd^{(0)}$ is not maximal in $\conv(\Pi)$.}
Then, let $\bmu \in \conv(\Pi)$ be such that $\bd^{(0)} \leq \bmu$ and $\bmu \neq \bd^{(0)}$.  
Then, $\bmu$ can be written as a convex combination of extreme points; 
i.e., $\bmu = p_1 \bd^{1} + \cdots + p_r \bd^{r}$
with $p_s > 0$ for all $s = 1, 2, \ldots, r$, $\sum_s p_s = 1$, 
$\bd^{s} \in \calE$ for all $s$ and the $\bd^{s}$ are all distinct. 
WLOG we may assume that $\bd^{s} \neq \bd^{(0)}$ for all $s$, 
since, for example, if $\bd^{1} = \bd^{(0)}$, 
then $\frac{\bmu - p_1 \bd^{1}}{1-p_1} = \frac{p_2}{1-p_1} \bd^{2} + \ldots \frac{p_r}{1-p_1} \bd^{r} \in \conv(\Pi)$, 
$\frac{\bmu - p_1 \bd^{1}}{1-p_1} \geq \bd^{(0)}$ and $\frac{\bmu - p_1 \bd^{1}}{1-p_1} \neq \bd^{(0)}$, 
so we can re-choose $\bmu$ to be $\frac{\bmu - p_1 \bd^{1}}{1-p_1}$.

Consider the set $\Pi' = \{\bd' \in \zp^N : \bd' \leq \bd^{s}~\text{for some}~s = 1, 2, \ldots, r\}$. 
By Assumption \ref{asmp:monotone}, $\Pi' \subseteq \Pi$. 
Since $\bd^{(0)} \leq \bmu = p_1 \bd^{1} + \cdots + p_r \bd^{r}$, 
it is easy to see that we can find $p'_{\bd'} \geq 0$ for $\bd' \in \Pi'$ such that 
$\sum_{\bd' \in \Pi'} p'_{\bd'} = 1$ and $\sum_{\bd'\in \Pi'} p'_{\bd'} \bd' = \bd^{(0)}$. 
Now $\bd^{(0)}$ is an extreme point, so we must have $p'_{\bd'} > 0 \Rightarrow \bd' = \bd^{(0)}$. 
Consider any $\bd' \in \Pi'$ such that $p'_{\bd'} > 0$. By definition, $\bd' \leq \bd^{s}$ for some $s$. 
Thus, $\bd^{(0)} = \bd' \leq \bd^s$. However, $\bd^s \neq \bd^{(0)}$ by assumption, 
which contradicts the maximality of $\bd^{(0)}$ in $\calE$. This establishes the lemma. 
\qed

\begin{lemma}\label{lem:max_extreme1}
Consider a schedule set $\Pi$ that satisfies Assumption \ref{asmp:monotone}, 
and let $\calE$ be the set of extreme points of $\conv(\Pi)$. 
Then, Assumption \ref{asmp:non-degeneracy2} holds if and only if 
$\conv(\Pi)$ has two distinct extreme points that are maximal.
\end{lemma}
\bpf
{Suppose that Assumption \ref{asmp:non-degeneracy2} does not hold. 
Then, the schedule set $\Pi$ has a unique maximal point, which dominates 
all other elements of $\Pi$. Thus, this unique maximal point dominates 
all points in $\conv(\Pi)$ as well, making it the unique maximal point in $\conv(\Pi)$.}
Now suppose that Assumption \ref{asmp:non-degeneracy2} holds. 
We will show that there are at least two distinct maximal extreme points in $\calE$. 
Suppose that $\calE$ has a unique maximal extreme point $\bd^{(0)}$. 
Then $\bd^{(0)}$ dominates all other extreme points in $\calE$, hence all points in $\conv(\calE)$ as well. 
But $\conv(\Pi) = \conv(\calE)$, so $\bd^{(0)}$ also dominates all points in $\conv(\Pi)$. 
This implies that $\Pi$ has a unique maximal element as well, contradicting Assumption \ref{asmp:non-degeneracy2}.
Thus, there are at least two distinct maximal extreme points in $\calE$, 
which, by Lemma \ref{lem:max_extreme}, are also maximal in $\conv(\Pi)$. \qed

\begin{lemma}\label{lem:communicating}
Fix any finite $k, v \in \zp$, and $\veps \in (0, 1)$. Consider some $\phi \in \Phi_k$ and $\psi \in \Psi_v$, 
and an $\veps$-majorizing channel $\calC = (\calX, \calY, C)$ that satisfies Assumption \ref{asmp:eps_maj}. 
Recall $\tilde{\Lambda}(\phi, \psi)$, the capacity region under the policy pair $(\phi, \psi)$, defined in \eqref{eq:stability_region}. 
For any $\blambda \in \tilde{\Lambda}(\phi, \psi)$, let $\pb_{\blambda}(\cdot)$ 
denote the stationary probability of the chain $\bW(\cdot)$. 
Then, there exists some $\delta > 0$ such that the following is true. 
For any $y \in \calY$ and $m \in \calM_r$, and for any $\blambda \in \tilde{\Lambda}(\phi, \psi)$, 
\begin{equation}\label{eq:reachability}
\pb_{\blambda}(Y(t) = y, M_r(t) = m) \geq \delta.
\end{equation}
\end{lemma}
\bpf Write $C = \veps C^0 + (1-\veps) C^1$ as in Eq. \eqref{eq:eps-dom-channel}, with $C^0$ having identical rows. 
Let $\delta' = \min\{C^0_{x,y} : x \in \calX, y \in \calY\}$. By Assumption \ref{asmp:eps_maj}, $\delta' > 0$.

Let $\blambda \in \tilde{\Lambda}(\phi, \psi)$, and consider any $y', y'' \in \calY$, and $m', m'' \in \calM_r$ with 
\begin{equation}
\pb_{\blambda}\lt(Y(t+1)=y', M_r(t+1) = m' \bbar Y(t)=y'', M_r(t) = m''\rt) > 0.
\label{eq:cond_lambda}
\end{equation}
{We claim that there exists $\tilde{\delta} > 0$ such that whenever \eqref{eq:cond_lambda} holds, then}
\begin{equation}
\pb_{\blambda}\lt(Y(t+1)=y', M_r(t+1) = m' \bbar Y(t)=y'', M_r(t) = m''\rt) \geq \tilde{\delta}.
\label{eq:cond_lambda2}
\end{equation}
To prove the claim, note that $M_r(t+1)$ is generated from $Y(t)$ and $M_r(t)$ by $\psi$ alone (see Eq. \eqref{eq:Mr_update}). 
Thus, $M_r(t+1)$ is conditionally independent from $Y(t+1)$, given $Y(t)$ and $M_r(t)$, 
and we can write
\begin{align}\label{eq:cond_indp}
&~\pb_{\blambda}\lt(Y(t+1)=y', M_r(t+1) = m' \bbar Y(t)=y'', M_r(t) = m''\rt) \nln
=&~\pb_{\blambda}\lt(Y(t+1)=y' \bbar Y(t)=y'', M_r(t) = m''\rt) \times \nln
& ~\times\pb_{\psi}\lt(M_r(t+1)=m' \bbar Y(t)=y'', M_r(t) = m''\rt), 
\end{align}
where the notation $\pb_{\psi}$ is used to emphasize the fact the conditional probability 
\begin{equation}
 \pb_{\psi}\lt(M_r(t+1)=m' \bbar Y(t)=y'', M_r(t) = m''\rt) \nnb
 \end{equation} depends only on the policy $\psi$, but not on $\blambda$. 
Since $C = \veps C^0 + (1-\veps) C^1$ and $\delta' = \min_{x, y} C^0_{x,y} > 0$, 
$\pb_{\blambda}\lt(Y(t+1)=y' \bbar Y(t)=y'', M_r(t) = m''\rt) \geq \veps \delta'$. 
Thus, 
\begin{align}
&~\pb_{\blambda}\lt(Y(t+1)=y', M_r(t+1) = m' \bbar Y(t)=y'', M_r(t) = m''\rt) \nln
\geq &~\veps \delta' \pb_{\psi}\lt(M_r(t+1)=m' \bbar Y(t)=y'', M_r(t) = m''\rt).
\end{align}
Also, by \eqref{eq:cond_lambda} and \eqref{eq:cond_indp}, 
we have 
\begin{equation}
\pb_{\psi}\lt(M_r(t+1)=m' \bbar Y(t)=y'', M_r(t) = m''\rt) > 0.
\end{equation}
The preceding conditional probability only depends on $\psi$, so we can find a uniform lower bound $\delta'' > 0$ 
with $\pb_{\psi}\lt(M_r(t+1)=m' \bbar Y(t)=y'', M_r(t) = m''\rt) \geq \delta''$. Therefore, we have 
\begin{equation}
\pb_{\blambda}\lt(Y(t+1)=y', M_r(t+1) = m' \bbar Y(t)=y'', M_r(t) = m''\rt) \geq \veps \delta' \delta''.
\end{equation}
By choosing $\tilde{\delta} = \veps \delta' \delta''$, we have established the claim. 

To prove the lemma, let $y_0, y \in \calY$ and $m_0, m \in \calM_r$. 
{By the irreducibility of $\bW(\cdot)$, there exists $T > 1$ such that 
\[
\pb_{\blambda}(Y(T) = y, M_r(T) = m \bbar Y(1) = y_0, M_r(1) = m_0) > 0.
\] 
Using Ineq. \eqref{eq:cond_lambda2} in the claim above, it is easy to show that 
\begin{equation}
\pb_{\blambda}(Y(T) = y, M_r(T) = m \bbar Y(1) = y_0, M_r(1) = m_0) \geq \tilde{\delta}^T.
\end{equation}
Since the set $\calY \times \calM_r$ is finite, we can choose $y_0 \in \calY$ and $m_0 \in \calM_r$ 
such that $\pb_{\blambda}(Y(1) = y_0, M_r(1) = m_0) > |\calY \times \calM_r|^{-1}$, 
and $T$ accordingly, so that 
\begin{equation}
\pb_{\blambda}(Y(1) = y, M_r(1) = m) = \pb_{\blambda}(Y(T) = y, M_r(T) = m) \geq \tilde{\delta}^T/|\calY \times \calM_r|.
\end{equation}
Letting $\delta = \min_{y \in \calY, m \in \calM} \tilde{\delta}^T/|\calY \times \calM_r|$, we have 
\begin{equation}
\pb_{\blambda}(Y(1) = y, M_r(1) = m) \geq \delta, 
\end{equation}
which establishes the lemma. \qed}

\emph{Proof of Theorem \ref{thm:kleps<1}.} 
We prove Theorem \ref{thm:kleps<1} by contradiction. 
Towards this end, 
suppose that there exists $\veps > 0$ and an $\veps$-majorizing channel $\calC = (\calX, \calY, C)$, 
such that $\rho^*_{k, v}(\calC) = 1$ for some $k \geq 0$ and $v \in \zp$. 

Let $\veps'>0$. Then there exist some $k' \in \zp$, and $(\phi, \psi) \in \Phi_{k'} \times \Psi_v$ 
such that 
\begin{equation}
\lt(1-\frac{\veps'}{2}\rt) \Lambda \subseteq \tilde{\Lambda}(\phi, \psi).
\end{equation}
By Lemma \ref{lem:max_extreme1}, let $\bd^{(1)}$ and $\bd^{(2)}$ be two distinct extreme points of $\conv(\Pi)$,  
both of which are maximal as well. 
Write $\blambda^{(j)} = (1-\veps') \bd^{(j)}$ for $j = 1, 2$. 
Then, $\blambda^{(j)} \in \tilde{\Lambda}(\phi, \psi)$ for both $j = 1, 2$. 
Consider $\pb_{\blambda^{(1)}}(\bD(t) = \bd^{(1)})$. 
By choosing $\veps'>0$ sufficiently small, we have $\pb_{\blambda^{(1)}}\lt(\bD(t) = \bd^{(1)}\rt) > 0$. 
Thus, there exist $y \in \calY$ and $m \in \calM_r$ such that 
\begin{equation}
\pb_{\blambda^{(1)}}\lt(\bD(t) = \bd^{(1)} \bbar Y(t) = y, M_r(t-1) = m\rt) > 0.
\end{equation}
Note that $\bD(t)$ is generated by the policy $\psi$, only based on $Y(t)$ and $M_r(t-1)$, 
so whenever $\blambda \in \tilde{\Lambda}(\phi, \psi)$, 
\begin{equation}
\pb_{\psi}\lt(\bD(t) = \bd^{(1)} \bbar Y(t) = y, M_r(t-1) = m\rt) = \pb_{\blambda}\lt(\bD(t) = \bd^{(1)} \bbar Y(t) = y, M_r(t-1) = m\rt).
\end{equation}
This implies that 
\begin{align}
&~\pb_{\psi}\lt(\bD(t) = \bd^{(1)} \bbar Y(t) = y, M_r(t-1) = m\rt) \nln
=&~\pb_{\blambda^{(1)}}\lt(\bD(t) = \bd^{(1)} \bbar Y(t) = y, M_r(t-1) = m\rt) \nln
=&~\pb_{\blambda^{(2)}}\lt(\bD(t) = \bd^{(1)} \bbar Y(t) = y, M_r(t-1) = m\rt).
\end{align}
Thus, 
\begin{align}
&~\pb_{\blambda^{(2)}} \lt(\bD(t) = \bd^{(1)}\rt) \nln
\geq &~\pb_{\blambda^{(2)}}\lt(\bD(t) = \bd^{(1)} \bbar Y(t) = y, M_r(t-1) = m\rt) \times 
\pb_{\blambda^{(2)}}\lt(Y(t) = y, M_r(t-1) = m\rt) \nln
= &~\pb_{\psi}\lt(\bD(t) = \bd^{(1)} \bbar Y(t) = y, M_r(t-1) = m\rt) \times \nln
&~\pb_{\blambda^{(2)}}\lt(Y(t) = y \bbar Y(t-1) = y, M_r(t-1) = m\rt) \times \pb_{\blambda^{(2)}}\lt(Y(t-1) = y, M_r(t-1) = m\rt) \nln
\geq &~\pb_{\psi}\lt(\bD(t) = \bd^{(1)} \bbar Y(t) = y, M_r(t-1) = m\rt) \veps \delta' \delta \bydef \delta_1, 
\label{eq:kleps<1_1}
\end{align}
where the last inequality uses the facts that (a) for an $\veps$-majorizing channel, 
$\pb (Y(t) = y \bbar Y(t-1) = y, M_r(t-1) = m) \geq \veps \delta'$ for any $y \in \calY$ 
(recall the definition of $\delta'$ in the proof of Lemma \ref{lem:communicating}), 
independent of the conditioning event in the earlier time slot, 
and $\pb_{\blambda^{(2)}}\lt(Y(t-1) = y, M_r(t-1) = m\rt) \geq \delta$ 
with $\delta$ in \eqref{eq:reachability}, by Lemma \ref{lem:communicating}. 

Ineq. \eqref{eq:kleps<1_1} implies that by choosing $\veps'$ sufficiently small, 
the policy pair $(\phi, \psi)$ cannot stabilize the system under the arrival rate vector $\blambda^{(2)}$. 
This contradicts the supposition that $\blambda^{(2)} \in \tilde{\Lambda}(\phi, \psi)$. 
Thus, for any $k', v \in \zp$, $\rho_{k', v}(\calC) < 1$.

To show that $\rho_{\infty, v}(\calC) < 1$ as well, simply note that the preceding argument in fact shows that 
there exists some $\delta_2 > 0$ such that $\rho_{k', v}(\calC) \leq 1-\delta_2$ for any $k', v \in \zp$. 
The theorem is hence established.
\qed

\subsection{Proof of Theorem \ref{thm:epsMajChar2}}
\label{app:thm:epsMajChar2}
\bpf  We begin by establishing the first statement of the theorem using a {lifting argument}. 
First, note that in Section \ref{sec:memoryless}, we will prove that $\rho^*_{k,0} = \rho^*_{0,0}$ 
for all $k\geq 0$ (see Theorem \ref{thm:stationaryCap}), so it suffices to show that 
\begin{equation}\label{eq:suff_epsMajChar2}
\rho^*_{0,0}(\calC) \leq 1-\veps(1-N^{-1})
\end{equation}
for any $\veps$-majorizing channel.  

Fix an $\veps$-majorizing channel, $\calC=(\calX,\calY,C)$, and write
\begin{equation}
C = \veps C^0 + (1-\veps)C^1,
\end{equation}
according to the decomposition in Definition \ref{def:eps-Major}. 
Use $\pb^C_x$ to denote the probability distribution over $\calY$ when the input signal is $x$, 
for the channel $C$; use the notation $\pb^{C^0}_x$ and $\pb^{C^1}_x$ in a similarly manner. 
Since $C^0$ has identical rows, we drop the subscript $x$ and use $\pb^{C^0}$ to denote a row of $C^0$. 
Then, for each $x$, the probability distribution $\pb^C_x$ is a mixture of $\pb^{C^0}$ and $\pb^{C^1}_x$. 
Thus, to generate the message $Y(t)$ from $X(t)$ through the channel $\calC$, 
it is equivalent to (a) first generate a ``switching'' Bernoulli random variable $\Xi(t)$, 
with $\pb(\Xi(t)=0) = \veps$, independent of everything else; 
(b) second, if $\Xi(t)=1$, $Y(t)$ is set to be $Y^1(t)$, the message generated from $X(t)$ through the channel 
$(\calX, \calY, C^1)$, and if $\Xi(t) = 0$, $Y(t)$ is set to be $Y^0(t)$, 
generated from the probability distribution $\pb^{C^0}$. 
See Figure \ref{fig:sysDiag_epsNoisy} for a pictorial illustration 
of this equivalent interpretation of the channel $\calC$.

\begin{figure}[ht]
\centering
\includegraphics[scale=.57]{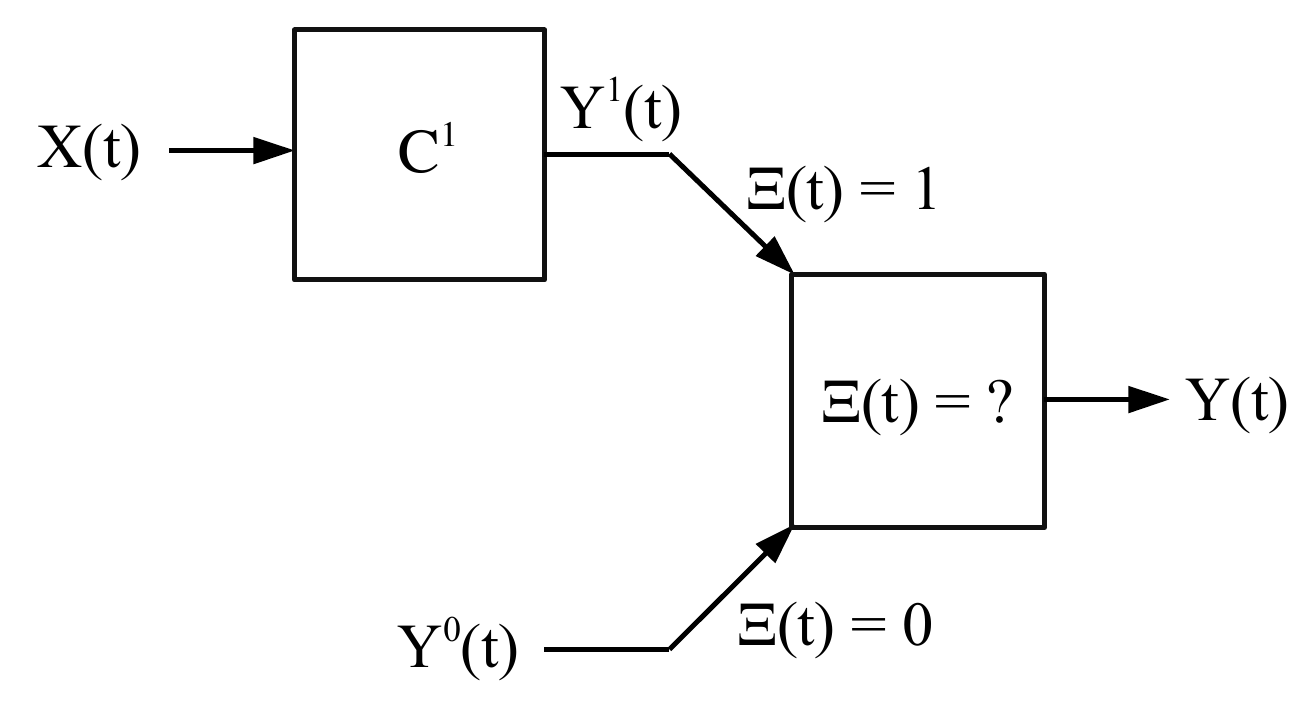}
\caption{Equivalent construction  of an $\veps$-majorizing channel.}
\label{fig:sysDiag_epsNoisy}
\end{figure}
We shall adopt the preceding interpretation of $\calC$ for the remainder of the proof. Consider a channel $\calC^+$ constructed by augmenting the output of $\calC$ {with} the value of the switching variable, $\Xi(t)$. Specifically, the output message of $\calC^+$ is equal to $(Y(t), \Xi(t))$, taking values in the alphabet set $\calY^+ = \calY\times \{0,1\}$. Note that $\calY^+$ is still a finite set and hence $\calC^+$ is a valid channel. 
It is easy to see that for any policy pair $(\phi, \psi)$ operating under the channel $\calC$, 
we can construct a corresponding pair $(\phi, \psi^+)$ operating under the channel $\calC^+$ 
by simply ignoring the variable $\Xi(t)$, so that $\tilde{\Lambda}(\phi, \psi) \subseteq \tilde{\Lambda}(\phi, \psi^+)$. 
We hence conclude that {$\rho^*_{0,0}(\calC)\leq \rho^*_{0,0}(\calC^+)$. Thus, to establish Ineq. \eqref{eq:suff_epsMajChar2},} 
it suffices to show that
\begin{equation}
\rho^*_{0,0}(\calC^+) \leq 1-\veps(1-N^{-1}). 
\label{eq:cal+eqv}
\end{equation}

Let $(\phi, \psi^+)$ be a pair of memoryless encoding and allocation policies that works with the channel $\calC^+$, 
and suppose that $(\phi, \psi^+)$ stabilizes the arrival rate vector $\blambda$.
Denote by $\bmu^0$ the $N$-dimensional vector where $\mu^0_i$ is the probability under $\psi^+$ that the server chooses to serve queue $i$, conditioning on $\Xi(t)=0$, i.e.,
\begin{equation}
\mu^0_i = \pb_{\psi^+}\lt(\bD(t) = \be^{(i)} \bbar \Xi(t)=0\rt), \quad t\in \zp.
\end{equation}
Note that conditioning on $\Xi(t)=0$, the message $Y(t)$ is equal to $Y^0(t)$, 
which is independent from the signal $X(t)$. 
Therefore, $\bmu^0$ does not change as the arrival rate vector $\blambda$ varies. 
Since the entries of $\bmu^0$ sum to no larger than $1$, it follows that
\begin{equation}
\min_{i=1,2,\ldots, N} \mu^0_i \leq \frac{1}{N}. 
\end{equation}
Let $\bmu = \bmu(\phi, \psi^+, \blambda)$ be the vector of stationary service rates 
defined in Eq.~\eqref{eq:stationary_service_rates} of Section \ref{ssec:station_rate}. 
Since $\pb(\Xi(t)=0)=\veps$, we conclude that the server will choose queue $i$ with probability at least $\mu^0_i \veps$. 
Fix $i^* \in \argmin_{i}\mu^0_i$. Then, 
\begin{equation}
\mu_{i^*} \leq 1- \veps \lt(\sum_{i\neq i^*} \mu^0_i\rt) \leq 1-\veps(1-N^{-1}).
\end{equation}
We claim that {the system cannot be stable} if $\blambda$ {admits} the following form: 
\begin{equation}\label{eq:unstable00}
\lambda_{i^*} > 1-\veps(1-N^{-1}) \quad \mbox{ and }\quad  \lambda_{i} = 0, \quad \forall i\neq i^*.
\end{equation}
for if \eqref{eq:unstable00} {were to hold}, then
\begin{equation}
\lambda_{i^*} > 1-\veps(1-N^{-1}) \geq \mu_{i^*},
\end{equation}
and queue $i^*$ {wouldn't} be stable.  Since $\phi$ and $\psi^+$ are arbitrary, we conclude that 
\begin{equation}
\rho^*_{0,0}(\calC^+) \leq 1-\veps(1-N^{-1}),
\end{equation}
which, by Eq.~\eqref{eq:cal+eqv}, proves the first statement of the theorem. 

We now prove the second statement of the theorem by considering the following example. 
Let $\calX  = \calY = {[N]}$, so that $\szX=\szY=N$. 
Define $I(t)$ to be the smallest index corresponding to a non-empty queue at time $t$, i.e., 
\begin{equation}
I(t) \bydef \min \{{i \in [N]}: Q_i(t) >0\}, 
\end{equation}
and let the encoder simply send the signal $I(t)$ at time $t$, i.e., $X(t) = I(t)$.
Consider the $N \times N$ channel matrix where
$C_{x,x} = 1-\veps(1-N^{-1})$ for all $x\in [N]$, 
and $C_{x,y} = \veps/N$ for all $x\neq y$. It is easy to verify that $C$ is in fact $\veps$-majorizing, where its  corresponding $C^0$ has all entries equal to $1/N$, and $C^1$ is the $N$-by-$N$ identity matrix. 
Finally, the allocation policy is simply to choose the queue whose index is equal to the message: 
$\bD(t) = \be^{(Y(t))}$ for all $t$.

Let $\blambda \in (1-\veps(1-N^{-1}))\Lambda$, then $\sum_{i=1}^N \lambda_i <  (1-\veps(1-N^{-1}))$. 
Consider the aggregate queue length process $\|\bQ(t)\|_1 \bydef \sum_{i=1}^N Q_i(t)$. Note that with probability $1-\veps(1-N^{-1})$, $Y(t)=X(t)$, in which case exactly $1$ job would depart from the system if and only if $\|\bQ(t)\|_1>0$. Otherwise, $Y(t) \neq X(t)$, and a job may or may not depart, depending on whether the chosen queue is empty or not. Therefore, the evolution of the process $\|\bQ(t)\|_1$ is stochastically dominated by the queue length process in a single-server-single-queue system with i.i.d.~arrivals, where the number of arrivals at time $t$ is equal to $\sum_{i=1}^N A_i(t)$, and the number of jobs to depart from the queue at time $t$ is a Bernoulli random variable with mean $1-\veps(1-N^{-1})$, if the queue is non-empty, and zero, otherwise. 
A simple Lyapunov function argument would show that in this system with arrival rate $\sum_{i=1}^N \lambda_i < 1-\veps(1-N^{-1})$ 
and service rate $1-\veps(1-N^{-1})$ when the queue is non-empty, the queue length process is positive recurrent. 
This shows that the channel $\calC = (\calX, \calY, C)$ satisfies
\begin{equation}
\rho^*_{0,0}(\calC) \geq 1-\veps(1-N^{-1}), 
\end{equation}
which, in light of the first claim of the theorem, implies that $\rho^*_{0,0}(\calC)$ is in fact equal to $1-\veps(1-N^{-1})$. This concludes the proof of Theorem \ref{thm:epsMajChar2}.



\subsection{Proof of Theorem \ref{thm:epsMajChar}}
\label{app:thm:epsMajChar}
In this section, we prove Theorem \ref{thm:epsMajChar}. 
First, we proceed with the proof of the first statement. 
Let $\veps \in (0, 1)$, and let $\calC = (\calX, \calY, C)$ be an $\veps$-majorizing channel. 
Let $\calC^+ = (\calX, \calY^+, C^+)$ be the augmented channel, which we used at the beginning of 
the proof of Theorem \ref{thm:epsMajChar2}. 
We will provide a constructive characterization of the constant $\rho(\veps, \Pi)$, 
and similar to Theorem \ref{thm:epsMajChar2}, 
it will be sufficient to show that 
\begin{equation}
\rho^*_{0,0}(\calC^+) \leq \rho(\veps, \Pi). 
\label{eq:cal+eqv2}
\end{equation}
Let $(\phi, \psi^+)$ be a pair of memoryless encoding and allocation policies. 
Also let $p(\bd) = \pb_{\psi^+} (\bD(t) = \bd \bbar \Xi(t) = 0)$ be the probability that 
the allocation policy chooses the schedule $\bd \in \Pi$ at time $t$, conditioning on $\Xi(t)=0$, 
and let $\bmu^0 = \sum_{\bd \in \Pi} p(\bd) \bd$ be the vector of stationary service rates, conditioning on $\Xi(t) = 0$. 
Note that the probabilities $p(\bd)$ are well-defined, independent of the arrival rate vector $\blambda$, 
since for any $y \in \calY$, the probabilities $\pb_{\psi^+} (\bD(t) = \bd \bbar Y(t) = y)$ depend only on the policy $\psi^+$, 
$\pb(Y(t) = y \bbar \Xi(t) = 0)$ depend only on the channel $\calC^+$, so 
\begin{equation}
p(\bd) = \pb_{\psi^+} (\bD(t) = \bd \bbar \Xi(t) = 0) 
= \sum_y \pb_{\psi^+} (\bD(t) = \bd \bbar Y(t) = y) \pb(Y(t) = y \bbar \Xi(t) = 0)
\end{equation}
does not depend on $\blambda$, for any $\bd \in \Pi$. 

Let $\Gamma$ be the set of achievable vectors of stationary service rates under the policy pair $(\phi, \psi^+)$. 
Then, it is easy to see that $\Gamma \subseteq (1-\veps)\conv(\Pi) + \veps \bmu^0$. 
We also let $\Gamma^+ = \{\blambda \in \rp^N : \blambda \leq (1-\veps)\conv(\Pi) + \veps \bmu^0\}$. 
Then, we must have $\tilde{\Lambda}(\phi, \psi^+) \subseteq \Gamma^+$. 
Finally, let $\rho(\veps, \bmu^0) = \sup \lt\{\rho > 0: \rho \Lambda \subseteq \Gamma^+ \rt\}$. 
Then, $\rho^*(\phi, \psi^+, \calC^+) \leq \rho(\veps, \bmu^0)$, 
where we recall the definition of $\rho^*(\phi, \psi^+, \calC^+)$ in Eq. \eqref{eq:cap_fac1}. 

We claim that $\rho(\veps, \bmu^0) < 1$ for all $\veps \in (0, 1)$ and $\bmu^0 \in \conv(\Pi)$. 
To prove the claim, we will show that 
\begin{itemize}
\item[(a)] $\rho(\veps, \bmu^0)$ is achievable, in the sense that 
$\rho(\veps, \bmu^0) \Lambda \subseteq \Gamma^+$; and
\item[(b)] $\Gamma^+\neq \conv(\Pi)$. 
\end{itemize}
To prove part (a), suppose, on the contrary, that $\rho(\veps, \bmu^0) \Lambda \subsetneq \Gamma^+$. 
Then, there exists $\bmu \in \Lambda$ such that $\rho(\veps, \bmu^0)\bmu \notin \Gamma^+$. 
Since $\Gamma^+$ is a compact set, its complement is open, 
and there exists $\delta > 0$ such that $\lt(\rho(\veps, \bmu^0) - \delta\rt) \bmu \notin \Gamma^+$. 
But this contradicts the definition of $\rho(\veps, \bmu^0)$. This proves part (a). 

To prove part (b), suppose on the contrary that $\Gamma^+ = \conv(\Pi)$. 
Then, $\Gamma^+$ has two distinct extreme points $\bd^{(1)}$ 
and $\bd^{(2)}$ that are also maximal. 
By definition, there exists $\bmu^1 \in \conv(\Pi)$ with $\bd^{(1)} \leq (1-\veps)\bmu^1 + \veps \bmu^0$. 
By the maximality of $\bd^{(1)}$, $\bd^{(1)} = (1-\veps)\bmu^1 + \veps \bmu^0$. 
Since $\bd^{(1)}$ is an extreme point, $\bd^{(1)} = \bmu^1 = \bmu^0$. 
Using a similar argument, we can show that $\bd^{(2)} = \bmu^0$. 
Thus, $\bd^{(1)} = \bd^{(2)}$, but this contradicts the supposition that 
$\bd^{(1)}$ and $\bd^{(2)}$ are distinct. This proves part (b). 

By part (a), since $\rho(\veps, \bmu^0) \Lambda \subseteq \Gamma^+$ and $\cl(\Lambda) = \conv(\Pi)$, 
we have $\rho(\veps, \bmu^0) \conv(\Pi) \subseteq \Gamma^+$.
Note also that $\Gamma^+ \subseteq \conv(\Pi)$. If $\rho(\veps, \bmu^0) = 1$, 
then $\conv(\Pi) \subseteq \Gamma^+ \subseteq \conv(\Pi)$, which implies that 
$\conv(\Pi) = \Gamma^+$, contradicting part (b). Thus, $\rho(\veps, \bmu^0) < 1$, which establishes the claim. 

Now, define 
\begin{equation}
\rho(\veps, \Pi) \bydef \sup_{\bmu^0 \in \conv(\Pi)} \rho(\veps, \bmu^0).
\end{equation}
It is not difficult to see that for any given $\veps \in (0, 1)$, 
$\rho(\veps, \bmu^0)$ is a continuous function of $\bmu^0$. 
$\conv(\Pi)$ is a compact set, so $\rho(\veps, \Pi) = \rho(\veps, \bmu^0)$ 
for some $\bmu^0 \in \conv(\Pi)$. 
This implies that $\rho(\veps, \Pi) = \rho(\veps, \bmu^0) < 1$, establishing the first statement of the theorem. 

To prove the second statement of the theorem, consider the following example. 
First, let $\bmu^0 \in \conv(\Pi)$ be such that $\rho(\veps, \Pi) = \rho(\veps, \bmu^0)$, 
and let $\lt(p(\bd)\rt)_{\bd \in \Pi}$ be a probability distribution with $\bmu^0 = \sum_{\bd \in \Pi} p(\bd) \bd$. 
We now turn to the construction of the channel $\calC = (\calX, \calY, C)$. 
Let $\calX = \calY = \Pi$. 
Let the channel matrix $C$ be given by: $C = \veps C^0 + (1-\veps) C^1$, 
where $C^0$ is the matrix with rows equal to $\lt(p(\bd)\rt)_{\bd \in \Pi}$, and $C^1$ is the identity matrix. 
The encoding policy $\phi$ is defined as follows. If $\bd \in \argmax_{\bd' \in \Pi} \langle \bQ(t-1), \bd' \rangle$, 
then $X(t) = \bd$. 
The allocation policy $\psi$ simply chooses the schedule $\bd$ that equals the received message $Y(t)$. 

Let $\blambda < (1-\veps) \conv(\Pi) + \veps \bmu^0$. Then, by a simple Lyapunov function argument, 
we can show that the policy pair $(\phi, \psi)$ stabilizes the arrival rate vector $\blambda$. 
This implies that the capacity factor $\rho^*(\phi, \psi, \calC) \geq \rho(\veps, \bmu^0) = \rho(\veps, \Pi)$. 
By the first statement, $\rho^*(\phi, \psi, \calC) \leq \rho(\veps, \Pi)$. 
Thus, $\rho^*(\phi, \psi, \calC) = \rho(\veps, \Pi)$, establishing the second statement of the theorem.
\qed

\subsection{Proof of Proposition \ref{prop:simple}}
\label{app:prop:simple}
\bpf
We will use the stationary distribution of the Markov chain $\bW(\cdot)$ 
under the policy pair $(\phi, \psi)$ to design a simple encoding policy $\phi^0 \in \Phi^S$, 
under which we will show that 
\begin{equation}
\bmu(\phi^0, \psi) = \bmu(\phi, \psi, \blambda) \geq \blambda.
\end{equation}
More specifically, consider the system under the policy pair $(\phi, \psi)$ and arrival rate vector $\blambda$. 
Because $\blambda\in \tilde{\Lambda}(\phi, \psi)$ by assumption, 
the Markov chain $\bW(\cdot)$ is positive recurrent, 
so it has a unique stationary distribution. 
Suppose that the chain $\bW(\cdot)$ is initialized with this unique stationary distribution at time $0$, 
and we use $\pb^{\infty}_{\phi, \psi}(\cdot)$ to denote the corresponding probability, 
where the subscripts $\phi, \psi$ are used to highlight the encoding and allocation policies employed. 

We now construct the simply encoding policy $\phi^0$ for $\psi$. 
In each time slot $t$, the encoder memory $M_e(t)$, is set to equal the current signal $X(t)$, i.e.,
\begin{equation}
M_e(t) = X(t) \quad \mbox{under $\phi^0$}, 
\end{equation}
and the signals $X(\cdot)$ under $\psi^0$ are generated according to the conditional probabilities
\begin{equation}
\pb \lt( X(t+1) = x' \bbar X(t)=x,  M_r(t) =m \rt) = \pb^{\infty}_{\phi,\psi}\lt (X(1)=x  \bbar   X(0)=x , M_r(0)=m\rt), 
\label{eq:psi0prob}
\end{equation}
for all $x \in \calX$ and $m\in \calM$. 
That is, the signal $X(t)$ will be sampled with respect to the stationary marginal probabilities 
of $X(1)$ conditioned on $(X(0),M_r(0))$, under $(\phi,\psi)$ and $\blambda$. 
Denote by $r(x,m)$ the stationary marginal probability
\begin{equation}
r(x,m)=\pb_{\phi, \psi}^{\infty}(X(0)=x, M_r(0)=m), \quad x\in \calX, m\in \calM_r. 
\end{equation}
Since $\bW(\cdot)$ is by assumption irreducible, 
$r(x,m) > 0$ for all $x$ and $m$. 
Thus, the conditional probabilities on the right-hand side in Eq.~\eqref{eq:psi0prob} are always well-defined.

Next, we show that under $(\phi^0, \psi)$, $(X(\cdot), M_r(\cdot))$ forms a homogeneous irreducible Markov chain. 
It is easy to check that $(X(\cdot), M_r(\cdot))$ is a homogeneous Markov chain, 
and we now show that it is irreducible. 
Define by $\calW$ to be the product space corresponding to 
the state space of $W(\cdot)$: $\calW\bydef  \calQ \times  \calM_e  \times \calX \times \calM_r$, 
and use $\pb_{\phi^0, \psi}$ to denote probabilities under $(\phi^0, \psi)$.
Fix $x,x'\in \calX$ and $m,m' \in \calM$. 
We have that
\begin{align}
&~\pb_{\phi^0, \psi}\lt( X(1) = x', M_r(1)=m' \bbar X(0)=x, M_r(0)=m\rt) \nln
=&~\pb^{\infty}_{\phi, \psi}\lt( X(1)=x', M_r(1)=m' \bbar X(0)=x, M_r(0)=m\rt). \label{eq:simple0}
\end{align}
Eq. \eqref{eq:simple0} can be derived as follows. 
\begin{align}
& ~\pb^{\infty}_{\phi,\psi}(X(1)=x', M_r(1) = m' \bbar X(0)=x,  M_r(0)=m)
\nln
\sk{a}{=} & ~\pb^{\infty}_{\phi,\psi}(X(1)=x'  \bbar  X(0)=x, M_r(0)=m)
 \cdot  \pb^{\infty}_{\phi,\psi}(M_r(1) = m' \bbar X(0)=x, M_r(0)=m)   \nln
\sk{b}{=} &  ~\pb^{\infty}_{\phi,\psi}(X(1)=x'  \bbar  X(0)=x, M_r(0)=m) \cdot \pb_{\psi}(M_r(1) = m' \bbar X(0)=x, M_r(0)=m)
\label{eq:rxm-ss} \\
\sk{c}{=} &~\pb_{\phi^0,\psi}(X(1)=x'  \bbar   X(0)=x, M_r(0)=m) \cdot \pb_{\phi^0,\psi}(M_r(1) = m' \bbar X(0)=x,  M_r(0)=m) \nln
\sk{d}{=}&~\pb_{\phi^0, \psi}\lt( X(1) = x', M_r(1)=m' \bbar X(0)=x, M_r(0)=m\rt). \nonumber
\end{align}
Step $(a)$ follows from the fact that, under the pair $(\phi, \psi)$ and conditional on $X(0)$ and $M_r(0)$, the only randomness in generating the next receiver memory state $M_r(1)$ is from $U_r(0)$, 
so $X(1)$ is conditionally independent from $M_r(1)$; 
this conditional independence can also be seen from the dependencies of variables illustrated in Figure \ref{fig:bayesian}.
In Step $(b)$, we remove the superscript $\infty$ and the subscript $\phi$ 
in the second term,  so as to emphasize the fact that the distribution of $M_r(1)$ is fully determined by the values of $X(0)$, $M_r(0)$ and the allocation policy $\psi$, regardless of how the overall chain $W(0)$ is initialized, or what encoding policy is used. 
Step $(c)$ follows from the definition of the policy $\phi^0$ in Eq.~\eqref{eq:psi0prob}. 
Finally, step $(d)$ is based on the same conditional independence property as that in Step $(a)$ in Eq.~\eqref{eq:rxm-ss}, which holds under any encoding and allocation policies  (Figure \ref{fig:bayesian}).

We proceed further to derive
\begin{align}
&~\pb_{\phi^0, \psi}\lt( X(1) = x', M_r(1)=m' \bbar X(0)=x, M_r(0)=m\rt) \nln
= &~\pb^{\infty}_{\phi, \psi}\lt( X(1)=x', M_r(1)=m' \bbar X(0)=x, M_r(0)=m\rt) \nln
= &~r(x,m)^{-1}\pb^{\infty}_{\phi, \psi}\lt( X(1)=x', M_r(1)=m',  X(0)=x, M_r(0)=m\rt) \nln
=  &~r(x,m)^{-1}\sum_{\stackrel
{\bw^1,\bw^0 \in \calW:}{x^1=x',m^1_r=m', x^0=x,m^0_r=m}} \pb^{\infty}_{\phi, \psi}\lt(\bW(1)=\bw^1, \bW(0)=\bw^0\rt) \nln
\sk{e}{\geq} & \sum_{\stackrel
{\bw^1,\bw^0 \in \calW:}{x^1=x',m^1_r=m', x^0=x,m^0_r=m}} \pb^{\infty}_{\phi, \psi}\lt(\bW(1)=\bw^1, \bW(0)=\bw^0\rt) \nln
=  & \sum_{\stackrel
{\bw^1,\bw^0 \in \calW:}{x^1=x',m^1_r=m', x^0=x,m^0_r=m}} \pb_{\phi, \psi}\lt(\bW(1)=\bw^1 \bbar \bW(0)=\bw^0\rt)\pb^{\infty}_{\phi,\psi}(\bW(0)=\bw^0),
\label{eq:onestepXM}
\end{align}
where we use the notation $\bw^j = (\bq^j, m^j_e, x^j, m^j_r)$, $j=0,1$. 
Here, step $(e)$ follows from the fact that $0<r(x,m)\leq 1$.
\begin{figure}[ht]
\centering
\includegraphics[scale=.32]{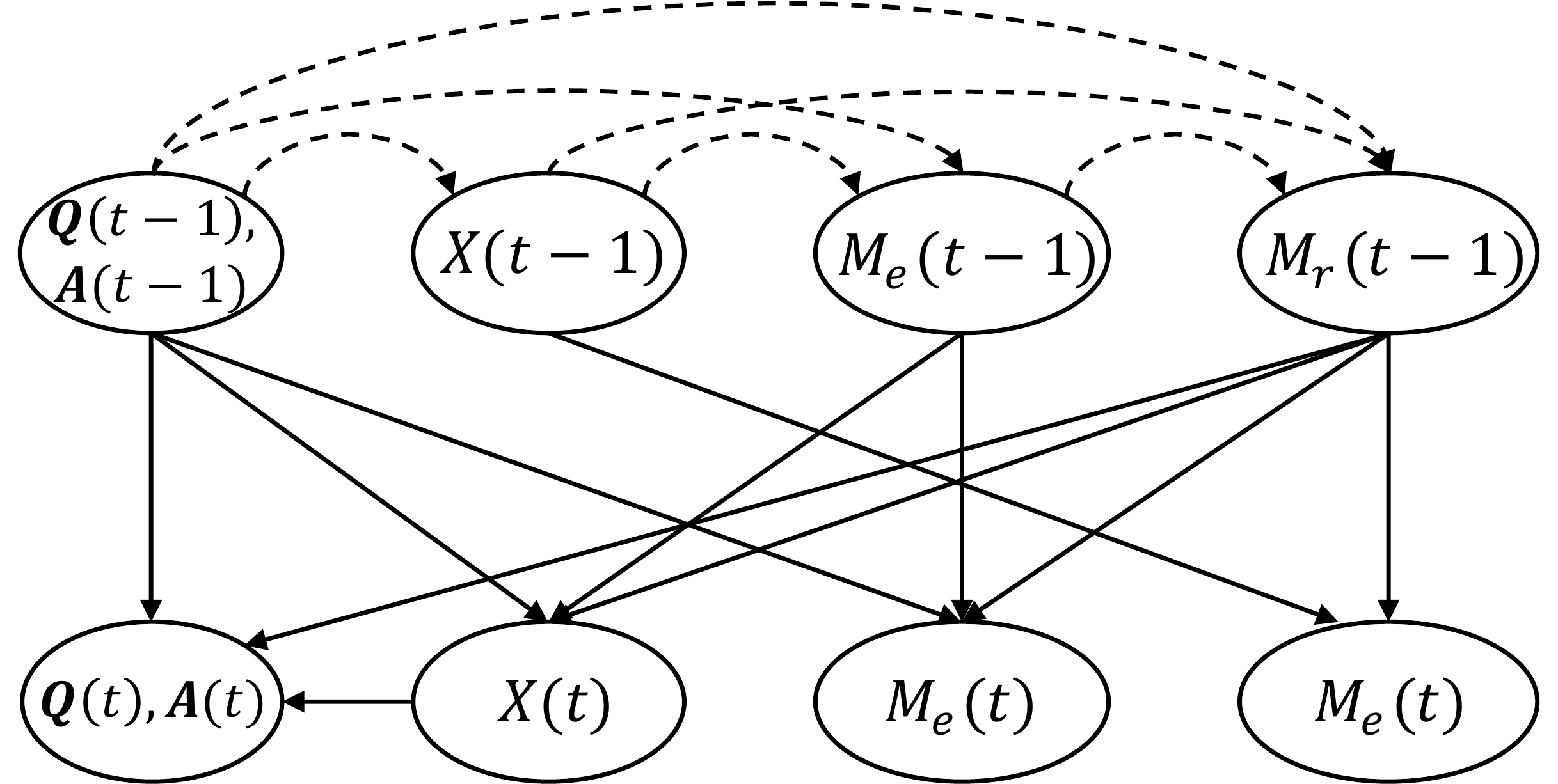}
\caption{A Bayesian network representation of the evolution of $\bW(\cdot)$. 
The directed lines represent dependence relations; 
for example, there is a directed line from $M_r(t-1)$ to $M_r(t)$, so $M_r(t)$ is dependent on $M_r(t-1)$. 
The set of dashed lines is a general representation of arbitrary dependence among 
$\bQ(t-1)$, $\bA(t-1)$, $X(t-1)$, $M_e(t-1)$ and $M_r(t-1)$, whereas solid lines indicate dependence relations 
from the policy update equations \eqref{eq:encoding}, \eqref{eq:D_update} and \eqref{eq:Mr_update}.}
\label{fig:bayesian}
\end{figure}

The irreducibility of $\bW(\cdot)$ implies that $\pb^{\infty}_{\phi,\psi}(\bW(0)=\bw^0)>0$ for all $\bw^0\in \calW$. Eq.~\eqref{eq:onestepXM} thus shows that  the one-step transition probability 
\begin{equation}
\pb_{\phi^0, \psi}\lt( X(1) = x', M_r(1)=m' \bbar X(0)=x, M_r(0)=m\rt)>0
\end{equation}
if any only if there exist two states $\bw^1,\bw^0 \in \calW$, with  $x^1=x',m^1_r=m', x^0=x,m^0_r=m$, such that
\begin{equation}
  \pb_{\phi, \psi}(\bW(1) =\bw^1 \bbar \bW(0)=\bw^0) >0.
 \end{equation}
Because $(X(\cdot),M_r(\cdot))$ is a time-homogenous Markov chain under $(\phi^0, \psi$), it is not difficult to extend the same observation from Eq.~\eqref{eq:onestepXM} to over multiple time slots, and conclude that, for any fixed $T>0$, 
\begin{equation}
 \pb_{\phi^0, \psi}\lt( X(T) = x', M_r(T)=m' \bbar X(0)=x, M_r(0)=m\rt)>0
 \end{equation} if and only if there exist $\bw^1,\bw^0 \in \calW$, with $x^1=x',m^1_r=m', x^0=x,m^0_r=m$, such that
 \begin{equation}
 \pb_{\phi, \psi}(\bW(T) =\bw^1 \bbar \bW(0)=\bw^0) >0.
 \end{equation}
Since the chain $\bW(\cdot)$ under $(\phi, \psi)$ is irreducible, $(X(\cdot), M_r(\cdot))$ under $(\phi^0, \psi)$ is also irreducible. 

We now proceed to show that
\[
\bmu(\phi^0, \psi) = \bmu(\phi, \psi, \blambda).
\]
By the irreducibility of the finite-state-space chain $(X(\cdot), M_r(\cdot))$ under $(\phi^0, \psi)$, 
it has a unique stationary distribution, which we denote by $(\bar{r}(x, m))_{x\in \calX, m \in \calM_r}$: 
\begin{equation}
\bar{r}(x,m)= \pb_{\phi^0, \psi}^{\infty} (X(0)=x, M_r(0)=m), \quad x\in \calX, m\in \calM_r. 
\end{equation}
Here $\pb_{\phi^0, \psi}^{\infty}$ denotes the probability associated with this stationary distribution. 
By definition, for each $x' \in \calX$ and $m' \in \calM_r$, $\bar{r}(x', m')$ satisfies the balance equation
\begin{equation}\label{eq:balance0}
\bar{r}(x',m') = \sum_{x \in \calX, m \in \calM_r} \pb_{\phi^0,\psi}(X(1)=x', M_r(1) = m' \bbar X(0)=x,  M_r(0)=m) \bar{r}(x,m). 
\end{equation}
By definition, we also have
\begin{equation}\label{eq:balance1}
r(x',m') = \sum_{x \in \calX, m \in \calM_r} \pb^{\infty}_{\phi,\psi}(X(1)=x', M_r(1) = m' \bbar X(0)=x,  M_r(0)=m) r(x,m). 
\end{equation}
Thus, by \eqref{eq:simple0}, $r(x', m')$ also satisfies the balance equation \eqref{eq:balance0}, 
and since $(\bar{r}(x, m))_{x, m}$ is unique, we must have
\begin{equation}
\bar r(x,m) = r(x,m), \quad \forall x \in \calX, m\in \calM_r. 
\label{eq:sameRrate}
\end{equation}
This implies that $\bmu(\phi^0, \psi) = \bmu(\phi, \psi, \blambda)$. 
By Lemma \ref{lem:stationary_service_rates} and the positive recurrence of $\bW(\cdot)$ under $(\phi, \psi)$ and $\blambda$, 
We have $\bmu(\phi, \psi, \blambda) \geq \blambda$. 
Therefore, $\bmu(\phi^0, \psi) \geq \blambda$ and this completes the proof of Proposition \ref{prop:simple}.  
\qed

\subsection{Proof of Lemma \ref{lem:epsCapkl}}
\label{app:lem:epsCapkl}
\bpf 
By the definition of $\rho^*_{k,v}$, for all $\veps \in (0, \rho^*_{k,v})$, there exist $\phi \in \Phi_k$ and $\psi \in \Psi_l$, such that 
\begin{equation}
\lt(\rho^*_{k,v}-\frac{\veps}{3}\rt) \Lambda \subseteq \tilde{\Lambda}(\phi, \psi). 
\label{eq:epsCapkl}
\end{equation}
By Ineq. \eqref{eq:epsCapkl} and Proposition \ref{prop:simple}, we have that for all $i\in [|\calF|]$, there exists $\phi_i \in \Pi^S$, such that
\begin{equation}\label{eq:perturbed0}
\bmu(\phi_i, \psi) \geq \lt(\rho^*_{k,v}-\frac{\veps}{3}\rt) \bmu^\calF_{i}, 
\end{equation}
and the chain $\lt(X(\cdot), M_r(\cdot)\rt)$ is irreducible under $(\phi_i, \psi)$. 

Let $\delta>0$, and consider perturbations $\psi^\delta$ and $\phi^\delta_i$ 
of the policies $\psi$ and $\phi_i$ for $i \in [|\calF|]$. 
Given the current signal $X(t)$ and receiver memory $M_r(t)$, 
$\psi^\delta$ updates $M_r(t+1) = M_r(t)$ with probability $\delta>0$, 
and it updates $M_r(t+1)$ according to $\psi$ with probability $1-\delta$; 
independently, $\phi^\delta_i$ updates $X(t+1) = X(t)$ with probability $\delta$, 
and updates $X(t+1)$ according to $\phi_i$ with probability $1-\delta$. 
Then, $(X(\cdot), M_r(\cdot))$ is clearly aperiodic under $(\phi^\delta_i, \psi^\delta)$. 
Furthermore, by choosing $\delta = \delta(\veps)$ sufficiently small, 
we can also make $(X(\cdot), M_r(\cdot))$ irreducible under $\lt(\phi^{\delta(\veps)}_i, \psi^{\delta(\veps)} \rt)$, 
and have
\begin{equation}\label{eq:perturbed1}
\bmu\lt(\phi^{\delta(\veps)}_i, \psi^{\delta(\veps)}\rt)  \geq \frac{\rho^*_{k,v}-2\veps/3}{\rho^*_{k,v}-\veps/3} \cdot\bmu(\phi_i, \psi). 
\end{equation}
Combining Ineqs. \eqref{eq:perturbed0} and \eqref{eq:perturbed1}, we have 
\begin{equation}\label{eq:perturbed2}
\bmu\lt(\phi^{\delta(\veps)}_i, \psi^{\delta(\veps)}\rt)  \geq \lt(\rho^*_{k,v}-\frac{2}{3}\veps\rt) \cdot\bmu(\phi_i, \psi). 
\end{equation}
With a slight abuse of notation, we write $\phi^\veps_i$ for $\phi^{\delta(\veps)}_i$, $i \in [|\calF|]$, 
and $\psi^\veps$ for $\psi^{\delta(\veps)}$. 
Then, Ineq. \eqref{eq:perturbed2} implies that 
$(\rho^*_{k,v}-\veps) \Lambda \subseteq \pconv\lt(\bmu(\Phi^\veps, \psi^\veps) \rt)$. 
\qed

\section{Infinite-Memory Receiver}
\label{app:sec:infRecMem}

In this section, we consider the case where the size of the receiver memory is infinite, i.e., $v = \infty$. 
The main result of this section is Theorem \ref{thm:infinite_rm} {(Item 3 of Theorem \ref{thm:main})}, 
which states that $\rho^*_{0, \infty}(\calC)=1$, 
for any informative channel $\calC$. In other words, we do not need to maintain any encoder memory 
in order to recover the maximal capacity region, when the receiver is equipped with infinite memory.

\begin{theorem}
\label{thm:infinite_rm}
 $\rho_{0, \infty}^*(\calC) = 1. $
\end{theorem}
The rest of this section is organized as follows. In Section \ref{ssec:EGL}, 
we provide a precise description of the encoding-allocation policy pair 
-- {\em Episodic Greedy Learning} (EGL) -- that will be used 
to prove Theorem \ref{thm:infinite_rm}. 
Then, in Section \ref{ssec:proof_infinite_rm}, 
we prove the main result proper. 
Finally, Appendix \ref{app:no_feedback} contains some discussion on 
a simple modification to EGL to the case of no memory-feedback, 
where we can show that $\rho^*_{\log N, \infty} = 1$, 
when the encoder does not have memory-feedback from the receiver. 

\subsection{Episodic Greedy Learning}\label{ssec:EGL}
To prove Theorem \ref{thm:infinite_rm}, 
we first describe the encoding-allocation policy pair $(\phi, \psi)$ that will be used, 
which we call {\em Episodic Greedy Learning} (EGL), 
where $\phi \in \Phi_0$ and $\psi \in \Psi_{\infty}$.
At a high level, the encoding policy $\phi$ is designed in such a way that 
the encoded signals are only used for the receiver to {\em estimate} 
the arrival rate vector, from the corresponding messages received. 
The allocation policy $\psi$ operates in {\em episodes} 
-- these are fixed-length blocks of time slots -- 
each one of which consists of two phases: 
\begin{enumerate}
\item Phase 1: {\em Learning}. The allocation policy learns an estimator of the arrival rate vector. 
\item Phase 2: {\em Deployment}. The allocation policy chooses a randomized schedule in each time slot, 
whose expectation strictly dominates the estimator produced in Phase 1. 
\end{enumerate}
The idea behind the allocation policy is that {if} $(a)$ the length of the deployment phase is substantially longer than that of the learning phase, and $(b)$ the arrival rate estimator generated from the learning phase is reasonably accurate, then the amount of service dedicated to each queue should be greater than the arrivals during each episode, in expectation. 

Before we proceed to a precise description of EGL, 
let us provide some remarks on the use of memory-feedback under the infinite-receiver-memory setting. 
On the one hand, with memory-feedback, the encoder potentially has access to a lot of past system information from the receiver, 
so it may seem intuitive that the encoder need not be equipped with any memory of her own. 
On the other hand, as we will see in the policy description, the memory-feedback is only used to {\em synchronize} 
the encoder and the receiver, so that the encoder knows the exact time in each episode. 
In fact, with some simple changes to the policy pair described in this section, 
it is possible to show that without memory-feedback, $\rho^*_{\log N, \infty}(\calC)=1$; 
see Appendix \ref{app:no_feedback} for details.

We now describe the EGL policy pair in detail. Let $B\in \N$ be the length of an episode, and $B_1$ and $B-B_1$ be the lengths of the learning and deployment phases, respectively. 
We also assume that $B_1$ is divisible by $N$.
Note that, because the encoder has access to the receiver memory through memory-feedback, we may assume that by recording time in $M_r$, the encoder knows the exact time relative to the start of the episode.

\emph{Encoding Policy}. Recall that the channel $\calC$ is informative, which implies that we can find $x_1, x_2 \in \calX$ and $y_1 \in \calY$ such that
 \begin{equation}
q_1 \bydef  \pb(Y(1) =y_1   \bbar X(1) = x_1) <  \pb(Y(1) = y_1  \bbar X(1)= x_2) \bydef q_2.  
 \end{equation}
During each time slot, the encoding policy first observes the current time $t$ relative to the start of the episode. 
Then, for $i \in [N]$, if $i \equiv t \bmod N$, the encoder sets $X(t) = x_1$ if $A_i(t-1) = 1$, and sets $X(t) = x_2$ otherwise. 
Thus, the encoder observes the $N$ queues in a round-robin manner, based on which the signals are decided.
Note that the signal $X(t)$ only depends on $\bA(t-1)$ and $M_r(t)$ (via memory-feedback), and the policy does not require any encoder memory. 
Let us also note that even though we have specified the encoder's decisions for the entire duration of the episode, 
this is not necessary, since the allocation policy will not be using the outputs of the channel during the deployment phase. 
The decisions of the encoding policy for the entire episode are provided for concreteness, and also for ease of reference in Appendix \ref{app:no_feedback}, 
when we discuss the case of no memory-feedback.
 
%
\emph{Allocation policy}: To define the allocation policy, we first introduce some notation. 
For $\bx\in \rp^N$ and a closed convex set $\calX\subset \rp^N$, define $\mbox{proj}(\bx,\calX)$ to be 
{\em the scaled projection of $\bx$ to the outer boundary of $\calX$}:
\begin{equation}
\mbox{proj}(\bx,\calX) \bydef a\bx, 
\end{equation}
where $a = \sup \{\tilde{a} \in \rp: \tilde{a}\bx \in \calX \}$. 

\begin{enumerate}
\item \emph{Learning Phase.} The allocation policy generates an estimator for $\blambda$, denoted by $\what \blambda$, as follows. For each $i \in [N]$, denote by $\what{p}_i$ the empirical frequency that the symbol $y_1$ is observed, 
out of all time slots $t$ for which $i = t \bmod N$, i.e., 
\begin{equation}\label{eq:learn_arrival}
\what{p}_i \bydef \frac{1}{B_1/N}\sum_{t: t = i \bmod N} \mathbb{I}_{\{Y(t) = y_1\}}. 
\end{equation}
Then, set
\begin{equation}
\what{\lambda}_i  = \frac{q_2 - \what{p}_i }{q_2-q_1}, \quad i = 1, \ldots, N. 
\end{equation}
We do not need to specify details of the allocation decisions in this phase, 
since the primary function of this phase is to learn an estimator of the arrival rate vector.
\item \emph{Deployment Phase.} Let $\alpha>0$ be a parameter, 
and let 
\begin{equation}
 \what{\blambda}^+ \bydef \left(\max \{0, \what{\lambda}_1\}, \ldots, \max \{0, \what{\lambda}_N\}\right).
 \end{equation} 
Consider the vector $\mbox{proj}\lt(\what{\blambda}^+  + \alpha \bOne,  \cl(\Lambda) \rt)$, 
the scaled projection of $\what{\blambda}^++\alpha \bOne$ to the boundary of $\cl(\Lambda)$, 
where we recall that $\bOne$ is the vector of all ones, and $\cl(\Lambda)$ is the closure of $\Lambda$. 
Since $\mbox{proj}\lt(\what{\blambda}^+  + \alpha \bOne,  \cl(\Lambda) \rt)$ is on the outer boundary of $\cl(\Lambda)$, 
there exists a random schedule $\bD$, distributed over the set $\Pi$ of schedules, so that 
\begin{equation}\label{eq:random_schedule}
\E\lt[\bD \bbar \what{\blambda}^+ \rt]= \mbox{proj}\lt(\what{\blambda}^+  + \alpha\bOne,  \cl(\Lambda) \rt)\footnote{Note that in Eq. \eqref{eq:random_schedule}, we have written the expectation of $\bD$ 
as being conditioned on $\what{\blambda}^+$, which is itself random, 
and the distribution of $\bD$ depends on the realization of $\what{\blambda}^+$.}. 
\end{equation}
During the deployment phase,  the allocation vector $\bD(t)$ chosen by the allocation policy are i.i.d. samples 
of the random schedule $\bD$. 
\end{enumerate}
\emph{Memory Requirement}. As we explained earlier, the encoder does not require any memory, due to the presence of memory-feedback. The receiver memories are used in the following ways: 
\begin{enumerate}
\item \emph{Time-keeping}: $\log(B)$ bits used to keep track of the relative time past since the start of an episode. 
\item \emph{Learning}: $B_1$ bits are used to store the messages received during the learning phase, one bit for each of the $B_1$ times slots. 
\end{enumerate}
Thus, altogether the allocation policy uses $B_1 + \log(B)$ bits of memory.

\subsection{Proof of Theorem \ref{thm:infinite_rm}}\label{ssec:proof_infinite_rm}
 

\bpf 
The following simple fact will be used throughout the proof: 
since $\Lambda$ is a bounded set, there exists $K > 0$ such that for all $\blambda' \in \Lambda$, 
$\blambda \leq K \bOne$.
%

Let $\veps \in (0, 1)$. Consider the EGL policy pair $(\phi, \psi) \in \Phi_0 \times \Psi_{\infty}$ described in Section \ref{ssec:EGL}. 
We will find suitable values of $B_1$, $B$ and $\alpha$, 
for which the capacity factor $\rho^*(\phi, \psi, \calC) \geq 1-\veps$. 

Let the arrival rate vector be $\blambda \in (1-\veps) \Lambda$. 
By Assumption \ref{asmp:non-degeneracy}, 
for all $i \in [N]$, we have $\be^{(i)} \in \Pi$. 
This implies that 
\[
\frac{\veps}{N} \bOne = \veps \sum_{i=1}^N \frac{1}{N} \be^{(i)} \leq \veps \conv(\Pi).
\]
Recalling $\blambda \in (1-\veps)\Lambda$, and $\Lambda = \pconv(\Pi)$, 
we have $\blambda + \frac{\veps}{N} \bOne \in \Lambda$.

By construction, the estimator $\what{p}_i$ defined in \eqref{eq:learn_arrival} is the empirical average from $B_1/N$ i.i.d.~samples of a Bernoulli distribution with mean $\lambda_i q_1 + (1-\lambda_i)q_2$. 
By the law of large numbers and the union bound,
there exists $B_1^*$ such that 
\begin{equation}
\pb\lt( \| \what{\blambda}^+ - \blambda\|_\infty \leq \frac{\veps}{3N} \rt) \geq 1-\frac{\veps}{12 K N}, 
\quad \forall B_1\geq B_1^*. 
\label{eq:Bdelt}
\end{equation}
For the rest of the proof, set 
\begin{equation}\label{eq:BB1alpha_choice}
B_1 = B_1^* N,~~B = \frac{8KN}{\veps} \cdot B_1,~~\text{and}~~\alpha = \frac{2\veps}{3N}.
\end{equation}
Consider the event $E = \lt\{\| \what{\blambda}^+ - \blambda\|_\infty \leq \frac{\veps}{3N}\rt\}$. 
We claim that under event $E$, we have 
\begin{equation}\label{eq:Bdelt1}
\what{\blambda}^+ + \alpha \bOne \in \Lambda,~~
\what{\blambda}^+ + \alpha \bOne \geq \blambda + \frac{\veps}{3N}\bOne, ~~\text{and}~~
\mbox{proj}(\what{\blambda}^+ + \alpha \bOne, \cl(\Lambda)) \geq \what{\blambda}^+ + \alpha \bOne. 
\end{equation}
First, observe that under event $E$, $\what{\blambda}^+ \leq \blambda + \frac{\veps}{3N}\bOne$, 
so $\what{\blambda}^++\alpha \bOne \leq  \blambda + \frac{\veps }{N}\bOne$. 
But $\blambda + \frac{\veps }{N} \bOne \in \Lambda$, 
so $\what{\blambda}^++\alpha \bOne \in \Lambda$ as well.
Second, under event $E$ we have $\what{\blambda}^+ \geq \blambda - \frac{\veps}{3N}\bOne$, 
so $\what{\blambda}^+ + \alpha \bOne \geq \blambda + \frac{\veps}{3N}\bOne$.
Finally, for any $\blambda' \in \Lambda$, $\mbox{proj}(\blambda', \cl(\Lambda)) \geq \blambda'$, 
so if $\what{\blambda}^+ + \alpha \bOne \in \Lambda$, then 
$\mbox{proj}(\what{\blambda}^+ + \alpha \bOne, \cl(\Lambda)) \geq \what{\blambda}^+ + \alpha \bOne$. 
This proves the claim. 

Let us now consider the randomized schedule $\bD$ chosen during the deployment phase. 
By construction, 
$\E\lt[\bD \bbar \what{\blambda}^+ \rt]= \mbox{proj}\lt(\what{\blambda}^+  + \alpha\bOne,  \cl(\Lambda) \rt)$.
By Eqs. \eqref{eq:Bdelt} and \eqref{eq:Bdelt1}, we have that 
\begin{eqnarray}
&& \E\lt[\bD\rt]  \nln
&= & \E\lt[ \E\lt[\bD \bbar \what{\blambda}^+ \rt]\rt] 
= \E\lt[ \mbox{proj}\lt(\what{\blambda}^+  + \alpha\bOne,  \cl(\Lambda) \rt) \rt] \nln
& \geq & \pb\lt(\mbox{proj}(\what{\blambda}^+ + \alpha \bOne, \cl(\Lambda)) \geq \what{\blambda}^+ + \alpha \bOne, 
~\what{\blambda}^+ + \alpha \bOne \geq \blambda + \frac{\veps}{3N}\bOne\rt) \lt(\blambda + \frac{\veps}{3N}\bOne\rt) \nln
& \geq & \pb \lt(E\rt) \lt(\blambda + \frac{\veps }{3N}\bOne\rt) 
\geq \lt(1-\frac{\veps}{12 K N}\rt) \lt(\blambda + \frac{\veps}{3N}\bOne\rt) \nln
& \geq & \blambda + \frac{\veps}{3N}\bOne - \frac{\veps}{12 K N} \cdot K \bOne \nln
&=& \blambda + \frac{\veps}{4N}\bOne, \label{eq:boundD}
\end{eqnarray}
where, in the last inequality, we have used the fact that $\blambda + \frac{\veps}{3N}\bOne \in \Lambda$, 
so $\blambda + \frac{\veps}{3N}\bOne \leq K\bOne$.

Denote by $\bA[n]$  and $\bD[n]$ the vector representing the total number of jobs that arrive and the total amount of service dedicated during the $n$th episode, respectively. We have that for all $n\in \N$
\begin{equation}\label{eq:aggAD}
\E(\bA[n])= B \blambda,~~\text{and}~~\E[\bD[n]] = (B-B_1)\E[\bD].
\end{equation}
Using \eqref{eq:BB1alpha_choice}, \eqref{eq:boundD} and \eqref{eq:aggAD}, we have
\begin{eqnarray}
\E[\bD[n]] & \geq & \lt(1 - \frac{\veps}{8KN}\rt) B \lt(\blambda + \frac{\veps}{4N}\bOne\rt) \nln
&\geq & B \lt(\blambda + \frac{\veps}{4N}\bOne\rt) - \frac{\veps}{8KN} \cdot B K\bOne \nln
& = & B  \lt(\blambda + \frac{\veps}{8N}\bOne\rt), \label{eq:bound_aggD}
\end{eqnarray}
where in the last inequality we have used the fact that $\blambda + \frac{\veps}{4N}\bOne \in \Lambda$, 
so $\blambda + \frac{\veps}{4N}\bOne \leq K \bOne$.

Similar to the proof of Theorem \ref{thm:EMWregion}, 
we consider the Markov chain $\bW[\cdot]$ sampled at the beginnings of the episodes. 
Noting that $\lt\{\lt(\bA[n], \bD[n]\rt)\rt\}$ are i.i.d.~sequences, 
and using \eqref{eq:aggAD}, \eqref{eq:bound_aggD}, and Proposition \ref{prop:absMW}, 
the sampled chain $\bW[\cdot]$ is positive recurrent under the EGL pair $(\phi, \psi)$ 
with parameters given in \eqref{eq:BB1alpha_choice}, whenever $\blambda \in (1-\veps)\Lambda$. 
This implies that $\rho^*(\phi, \psi, \calC) \geq 1-\veps$. 
But $\veps \in (0, 1)$ is arbitrary, so we must have $\rho^*_{0, \infty}(\calC) = 1$. \qed 

%

\subsection{Infinite Receiver Memory: No Memory-Feedback}\label{app:no_feedback}
In this subsection, we consider the case where the encoder does not have memory-feedback 
from the receiver; in other words, other than the allocation vector $\bD(t)$ that the encoder can observe 
from the receiver in each time slot, the encoder does not have access to any content of the receiver memory. 
We still assume that the receiver is equipped with infinite memory. 
Then, we can prove the following
\begin{theorem}\label{thm:no_feedback}
Under no memory-feedback, $\rho^*_{\log N, \infty}(\calC) = 1$.
\end{theorem}
\emph{Proof Sketch.} The proof of the theorem uses a simple, modified version of EGL, 
and is then essentially that of Theorem \ref{thm:infinite_rm} verbatim. 
Hence, we only provide a proof sketch here. 

Recall the EGL policy pair described in Section \ref{ssec:EGL}. 
Consider the following changes to EGL: 
\begin{enumerate}
\item The encoding policy still observes the queues in a round-robin manner, 
and sends signals in the same way as EGL. However, instead of keeping track of the time 
relative to the start of the current episode, the encoder memory maintains 
the index of the queue currently observed. More specifically, 
if the current time is $t$ and $t \equiv i \bmod N$, then $M_e(t) = i$. 
The memory content is updated as $M_e(t+1) = M_e(t) + 1$ if $M_e(t) < N$, 
and $M_e(t+1) = 1$ if $M_e(t) = N$. 
Note that the encoder only needs $\log N$ bits of memory to keep track 
of the queue indices, and does not require memory-feedback from the receiver to know 
which queue to observe. 
\item At time $0$, the encoder and the receiver synchronize for the receiver 
to know that at time $1$, the encoder sends a signal based on the state of the first queue. 
The encoder and the receiver do not need to synchronize after time $0$.
\end{enumerate}
Under the preceding changes to the EGL policy pair, essentially the same argument 
from the proof of Theorem \ref{thm:infinite_rm} can be used to show that 
under no memory-feedback, $\rho^*_{\log N, \infty}(\calC) = 1$. 
One key point to note is that without the synchronization between the encoder and 
the receiver at time $0$, the receiver may only obtain estimates of the arrival rates 
up to a cyclic permutation on the queue indices. Synchronization resolves this issue, 
and lets the receiver know which estimate correspond to which queue. 
\qed

\section{{Finding the Capacity Factor and Optimal Policies} }
\label{app:calcRho}
We demonstrate in this section how to calculate the capacity factor $\rho^*_{\infty,v}(\calC)$ as the optimal value of a polynomial optimization problem over finite-dimensional matrices. Note that, by Theorem \ref{thm:main}, this will cover all $\rho^*_{k, v}(\calC)$ for $k\geq K(\Pi, \calX)$, and serve as an upper bound on $\rho^*_{k,v}$ for $k <  K(\Pi, \calX)$. Moreover, the optimal solutions to this optimization problem  lead to the allocation policy ($\psi^\epsilon)$ and simple encoding policies ($\Phi^\epsilon$) that will be used by the Episodic Max-Weight policy (Lemma \ref{lem:epsCapkl}, Section \ref{ssec:EMW}) to achieve the capacity factor. 

We first consider the case where $v \geq 1$ (i.e., the  receiver is not memoryless). Fix  $v \in \N$. The chain $\{(X(t), M_r(t), \bD(t))\}_{t\in \N}$ is Markov with the following transition dynamics. Let a  simple encoding policy (Definition \ref{def:simpPol}) be parameterized by the $(|\calM_r| |\calX|) \times |\calX|$ row-stochastic matrix, $G^E$, with 
\begin{equation}
G^E_{(m,x),x' } = \pb(X(t+1)=x'\bbar M_r(t)=m, X(t) = x). 
\end{equation}
Similarly, an allocation policy $\psi$ can be parameterized by the pair $(G^A, H^A)$, where $G^A$ is an $(|\calM_r|  |\calY|)\times |\calM_r|$ matrix, with
\begin{equation}
G^A_{(m,y), m'} = \pb(M_r(t+1) = m' \bbar M_r(t) = m, Y(t) = y), 
\end{equation}
and  $H^A$ is an $(|\calM_r|  |\calY|)\times |\Pi|$ matrix, with 
\begin{equation}
H^A_{(m,y), \bd} = \pb(\bD(t) = \bd \bbar  M_r(t) = m, Y(t) = y). 
\end{equation}

Denote by $G^S$ the transition matrix associated with the chain $\{(X(t), M_r(t), \bD(t))\}_{t\in \N}$: 
\begin{align}
& G^S_{(m,x,\bd), (m',x',\bd')}  \nln
=& \pb\big(X(t+1)=x', M_r(t+1)=m', \bD(t+1)=\bd' \nln
& \quad  \bbar M_r(t)=m, X(t) = x, \bD(t) = \bd \big). 
\end{align}
We can write $G^S$ as a function of $G^E$, $G^A$ and $G^E$: 
\begin{align}
 G^S_{(m,x,\bd), (m',x',\bd')}  = & G^E_{(m,x), x'}  \lt( \sum_{y \in \calY} C_{x',y} G^A_{(m,y), m'} H^A_{(m,y),\bd}\rt), 
\label{eq:gs}
\end{align}
where $C$ is the channel matrix, with $C_{x,y} = \pb(Y(t) = y \bbar X(t) =x)$. Note that $G^E, G^E$ and $H^A$ are row-stochastic matrices chosen by the system designer, while $C$ is given. To ensure that the resulting $G^S$ is irreducible, we may perturb the entries in $C$ by a very small amount so that all entries are positive, and similarly, we may constrain the entries of the row-stochastic matrices to be bounded from below by a small constant. The irreducibility of $G^S$  implies that it is associated with a unique stationary distribution, given by 
\begin{equation}
\bp = \lt(I - G^S\rt)^{-1}. 
\end{equation}
Denote by $\what{\bp}$ the marginalized stationary distribution over the allocation vectors: 
\begin{equation}
\what{\bp}_{\bd} = \sum_{m\in \calM_r, x\in \calX} \bp^S_{(m,x,\bd)}, \quad \bd \in \Pi. 
\end{equation}
The resulting stationary service rate is given by 
\begin{equation}
\bmu(G^E, (G^A, H^A)) = \sum_{\bd \in \Pi} \what{\bp}_{\bd} \cdot \bd. 
\end{equation}

Using the above construction, we now formulate the optimization problem  that will lead to the capacity factor. Recall from Lemma \ref{lem:epsCapkl} that the capacity factor is given by the minimal shrinkage to the maximum capacity region such that it can be dominated by the set of service rates achievable through simple encoding policies. We can therefore compute $\rho^*_{\infty, v}(\calC)$ as follows. Let $\calE'$ be the set of maximal schedules in $\calE$, where $\calE' = \{\bd^{(i)}\}_{i = 1, \ldots, |\calE'|}$.  Consider the following polynomial optimization problem:  
\begin{align}
\mbox{maximize} & \quad \rho \nln
\mbox{subject to} &\quad  \bmu\big(G^E(i), \big(G^A, H^A \big) \big) \geq  \rho\bd^{(i)}, \nln
& \quad  i = 1, \ldots, |\calE'|, 
\label{eq:optProbRhoEll}
\end{align}
where the variables to be optimized are the row-stochastic matrices $G^A$, $H^A$, and  $\big\{G^E(i)\big\}_{i = 1, \ldots, |\calE'|}$. Denote by $\overline{\rho}$ and $\Big( \big\{\overline{G}^E(i) \big\}_{i =1, \ldots, |\calE'|},$ $\big( \overline{G}^A, \overline{H}^A \big) \Big)$ the optimal value and an optimal solution of \eqref{eq:optProbRhoEll}, respectively. We have that the optimal value corresponds to the capacity factor: 
\begin{equation}
\rho^*_{\infty, v}(\calC) = \overline{\rho}. 
\end{equation}
Furthermore, to construct the Episodic Max Weight policy that achieves the capacity factor, the allocation policy ($\psi^\epsilon$ in Lemma \ref{lem:epsCapkl}) and the set of simple encoding policies ($\Phi^\epsilon$ in Lemma \ref{lem:epsCapkl})  are given by those associated with $\big( \overline{G}^A, \overline{H}^A \big)$ and $\big\{\overline{G}^E(i) \big\}_{i =1, \ldots, |\calE'|}$, respectively. 

\emph{Special case of memoryless receiver}. When  $v=0$, the optimization problem in \eqref{eq:optProbRhoEll} can be further simplified.  Recall the notation of rate allocation matrix $\Theta$ in  Eq.~\eqref{eq:psi_matrix}, and schedule matrix $S$ in \eqref{eq:schedule_matrix}. The capacity factor  $\rho^*_{\infty,0}(\calC)$ is the optimal value of the following optimization problem: 
\begin{align}
\mbox{maximize} & \quad \rho \nln
\mbox{subject to} &\quad \br^{(i)}C \Theta S \geq \rho \bd^{(i)}, \quad i =1, \ldots, |\calE'|, \nln
& \quad  \br^{(i)} \geq 0, \, \sum_{x \in \calX}  \br^{(i)}_{x} = 1, \quad i =1, \ldots, |\calE'|, 
\label{eq:optProbRho0}
\end{align}
where the variables to be optimized are the probability vectors $\{\br^{(i)}\}_{ i =1, \ldots, |\calE'|}$  and row-stochastic matrix $\Theta$. Here, $\br^{(i)}$ represents the probabilities over the set of input symbols, $\calX$. The allocation policy that achieves the capacity factor corresponds to the matrix $\Theta^*$ in an optimal solution of  \eqref{eq:optProbRho0}. 

Notably, enabled by our theoretical results, the above derivations show that $\rho^*_{\infty, v}(\calC)$ can be calculated by solving polynomial optimization problems over \emph{finite-dimensional} matrices. In contrast, without Theorem \ref{thm:main}, it would have not been clear \emph{a priori} how to compute $\rho^*_{\infty, v}(\calC)$, since it is defined as the supremum over an unbounded family of encoding policies, who may take as input the entire queue lengths, as well as an encoding memory state ($M_e(t)$) with an unbounded size. In the same way, identifying the optimal encoding and allocation policies that achieve the capacity factor would have been quite difficult if one were to solve it via brute force.  Admittedly, the optimization problems in \eqref{eq:optProbRhoEll} and \eqref{eq:optProbRho0} still have their drawbacks: they may be non-convex and could scale poorly as the size of the receiver memory, $v$, or that of the maximal schedules, $|\calE'|$,  becomes larges. It would be an interesting future direction to investigate how these optimization problems can be solved efficiently.

\section{Glossary of Frequently Used Symbols}
\vspace{5pt}
\begin{tabular}{l*{2}{l}r}
\hline
$\bA(t)$& arrival vector in time slot $t$\\
$\calC$, $C$ & channel, channel matrix\\
$\bD(t)$&  chosen allocation / schedule in time slot $t$\\
$k$ & encoder memory size (bits)\\
$v$ & receiver memory size (bits)\\
$\bLam$ & maximum capacity region\\ 
$M_e, M_r$ & encoder / receiver memory\\ 
$N$ & system size /  number of queues\\
$\Pi$ &  set of allowable allocations / schedules\\
$\bQ(t)$ & queue state in time slot $t$\\
$\rho^*_{k,v}(\calC)$ & $(k,v)$-capacity factor\\
$\Theta$ & rate allocation matrix for memoryless allocation policies\\
$X(t), \calX$ & input symbol in time slot $t$, input alphabet\\
$Y(t), \calY$ & output symbol in time slot $t$, output alphabet\\
\hline
\end{tabular}

\ifx \useplain\undefined
\end{APPENDICES}
\fi

\end{document}